\documentclass[journal, 11pt, onecolumn, draftclsnofoot]{IEEEtran}

\usepackage{eurosym}
\usepackage[cmex10]{amsmath}
\usepackage{amsthm}
\usepackage{amsfonts}
\usepackage{amssymb}
\usepackage{graphicx}
\usepackage{epstopdf}
\usepackage{algpseudocode}
\usepackage{algorithmicx}
\usepackage[caption=false,font=footnotesize]{subfig}
\usepackage[utf8]{inputenc}
\usepackage{longtable}
\usepackage[usenames,dvipsnames]{color}
\usepackage{colortbl}
\usepackage{multirow}
\usepackage{verbatim}
\usepackage{hyperref}
\usepackage{cite}
\usepackage{cases}

\theoremstyle{definition}
\newtheorem{rem}{Remark}
\theoremstyle{plain}
\newtheorem{pro}{Proposition}

\newtheorem{thm}{Theorem}

\begin{document}

\title{Consensus Labeled Random Finite Set Filtering\\for Distributed Multi-Object Tracking}
\author{Claudio Fantacci, Ba-Ngu Vo, Ba-Tuong Vo, Giorgio Battistelli and
Luigi Chisci\thanks{%
Claudio Fantacci was with the Dipartimento di Ingegneria dell'Informazione,
Universit\`a di Firenze, Florence, 50139, Italy, on leave at the Curtin
University, Bentley, WA, 6102, Australia, in the period January-July 2014.
He is currently with the Istituto Italiano di Tecnologia (IIT), Genoa, Italy
(e-mail: \href{mailto:claudio.fantacci@unifi.it}{claudio.fantacci@unifi.it}, \href{mailto:claudio.fantacci@iit.it}{claudio.fantacci@iit.it}).%
}\thanks{%
Ba-Ngu Vo and Ba-Tuong Vo are with the Department of Electrical and Computer
Engineering, Curtin University, Bentley, WA, 6102, Australia (e-mail: \href{mailto:ba-ngu.vo@curtin.edu.au}%
{ba-ngu.vo@curtin.edu.au}, \href{mailto:ba-tuong@curtin.edu.au}{%
ba-tuong@curtin.edu.au}).}\thanks{%
Giorgio Battistelli and Luigi Chisci are with the Dipartimento di Ingegneria
dell'Informazione, Universit\`a di Firenze, Florence, 50139, Italy (e-mail:
\href{mailto:giorgio.battistelli@unifi.it}{giorgio.battistelli@unifi.it},
\href{mailto:luigi.chisci@unifi.it}{luigi.chisci@unifi.it}).}}
\maketitle

\begin{abstract}
This paper addresses distributed multi-object tracking over a network of
heterogeneous and geographically dispersed nodes with sensing, communication
and processing capabilities. The main contribution is an approach to
distributed multi-object estimation based on labeled Random Finite Sets
(RFSs) and dynamic Bayesian inference, which enables the development of two
novel consensus tracking filters, namely a Consensus Marginalized $\delta $%
-Generalized Labeled Multi-Bernoulli and Consensus Labeled Multi-Bernoulli
tracking filter. The proposed algorithms provide fully distributed, scalable
and computationally efficient solutions for multi-object tracking.
Simulation experiments via Gaussian mixture implementations confirm the
effectiveness of the proposed approach on challenging scenarios.
\end{abstract}


\section{Introduction}

\label{sec:intro} \textit{Multi-Object Tracking} (MOT) involves the on-line
estimation of an unknown and time-varying number of objects and their
individual trajectories from sensor data \cite%
{farina1,farina2,barshalom1,barshalom2,Goodmanetal97,blackmanpopoli,Mahler07, VoMTT15}. In
a multiple object scenario, the sensor observations are affected by
misdetection (e.g., occlusions, low radar cross section, etc.) and false
alarms (e.g., observations from the environment, clutter, etc.), which is
further compounded by association uncertainty, i.e. it is not known which
object generated which measurement. The key challenges in MOT include
\textit{detection uncertainty}, \textit{clutter}, and \textit{data
association uncertainty}. Numerous multi-object tracking algorithms have
been developed in the literature and most of these fall under the three
major paradigms of \textit{Multiple Hypothesis Tracking} (MHT) \cite%
{reid,blackmanpopoli}, \textit{Joint Probabilistic Data Association} (JPDA)
\cite{barshalom2}, and \textit{Random Finite Set} (RFS) \cite{Mahler07}.

Recent advances in wireless sensor technology inspired the development of
large sensor networks consisting of radio-interconnected nodes (or agents)
with sensing, communication and processing capabilities \cite{SPM}. The main
goal of such a net-centric sensing paradigm is to provide a more complete
picture of the environment by combining information from many individual
nodes (usually with limited observability) using a suitable \textit{%
information fusion} procedure, in a way that is \textit{scalable} (with the
number of nodes), \textit{flexible} and \textit{reliable} (i.e. \textit{%
resilient} to failures) \cite{SPM}. Reaping the benefits of a sensor network
calls for distributed architectures and algorithms in which individual
agents can operate with neither central fusion node nor knowledge of the
information flow in the network \cite{Olfati}.

The wide applicability of MOT together with the emergence of net-centric
sensing motivate the investigation of \textit{Distributed Multi-Object
Tracking} (DMOT). {S}calability with respect to network size, lack of a
fusion center as well as knowledge of the network topology call for a
\textit{consensus} approach to achieve a collective information fusion over
the network \cite{Olfati,Xiao,Olfati2,Tomlin,Chiuso,Calafiore,Stankovic,Sayed,Farina,auto2014}. In fact, consensus has recently
emerged as a powerful tool for distributed computation over networks \cite%
{Xiao,Olfati}, including parameter/state estimation \cite{Olfati2,Tomlin,Chiuso,Calafiore,Stankovic,Sayed,Farina,auto2014}.
Furthermore, a robust (possibly
suboptimal) information fusion procedure is needed to combat the data
incest problem that causes \textit{double counting} of information.
To this end, \textit{Chernoff fusion} \cite{info,mori1}, also known as \textit{Generalized Covariance Intersection} \cite{gci,hurley}
 (that encompasses \textit{Covariance Intersection} \cite{ci,ciws}) or \textit{Kullback-Leibler average} \cite{auto2014,ccphd},
is adopted to fuse multi-object densities computed by various nodes of the network.
Furthermore, it was proven in \cite{ciws} for the single-object case, and subsequently in \cite{SPIE} for the multi-object case,
that Chernoff fusion is inherently immune to the double counting of information, thereby justifying
its use in a distributed setting wherein the nodes operate without knowledge about their common information.

While the challenges in MOT are further compounded in a distributed
architecture, the notion of \textit{multi-object probability density} in the
RFS formulation enables consensus for distributed state estimation to be
directly applied to multi-object systems
\cite{mahler-chap-2012,ccphd,SPIE,emd,ccjpda}.
Indeed, a robust and tractable
multi-object fusion solution based on Kullback-Leibler averaging, together
with the \textit{Consensus Cardinalized Probability Hypothesis Density}
(CPHD) filter have been proposed in \cite{ccphd}. However, this RFS-based
filtering solution does not provide estimates of the object trajectories and
suffers from the so-called \textquotedblleft \textit{spooky effect}%
\textquotedblright\ \cite{spooky}.
Note that one of the original intents of the RFS formulation is to propagate the distribution of the set of tracks via the use of labels, see \cite[p. 135, pp. 196-197]%
{Goodmanetal97}, \cite[p. 506]{Mahler07}. However, this capability was
overshadowed by the popularity of unlabeled RFS-based filters
such as PHD, CPHD, and multi-Bernoulli \cite{MahlerPHD2,MahlerCPHDAES,Mahler07,VVC09,VVPS10}.

This paper proposes the first consensus DMOT algorithms based on the
recently introduced labeled RFS formulation \cite{vovo1}. This formulation
admits a tractable analytical MOT solution called the \textit{$\delta $%
-Generalized Labeled Multi-Bernoulli} ($\delta $-GLMB) filter \cite{vovo2}
that does not suffer from the \textquotedblleft \textit{spooky effect}%
\textquotedblright , and more importantly, outputs trajectories of objects.
Furthermore, efficient approximations that preserve key summary statistics
such as the \textit{Marginalized $\delta $-GLMB} (M$\delta $-GLMB) and the
\textit{Labeled Multi-Bernoulli} (LMB) filters\ have also been developed
\cite{mdglmbf,lmbf}. In this paper, it is shown that the M$\delta $-GLMB and
LMB densities are algebraically closed under Kullback-Leibler averaging, and
novel consensus DMOT M$\delta $-GLMB and LMB filters are developed.

The rest of the paper is organized as follows. Section \ref{sec:back}
presents notation, the network model, and background on Bayesian filtering,
RFSs, and distributed estimation. Section \ref{sec:lmbnwgm} presents the
Kullback-Leibler average based fusion rules for M$\delta $-GLMB and LMB
densities. Section \ref{sec:clmbif} describes the multi-object Bayesian
recursion with labeled RFSs and presents the novel \textit{Consensus M$%
\delta $-GLMB} and \textit{Consensus LMB} filters with \textit{Gaussian
Mixture} (GM) implementation. Section \ref{sec:performance} provides a
performance evaluation of the proposed DMOT filters via simulated case
studies. Concluding remarks and perspectives for future work are given in
Section \ref{sec:conclusions}.

\section{Background and problem formulation}

\label{sec:back}

\subsection{Notation}

\label{ssec:notation} Throughout the paper, we use the standard inner
product notation $\left\langle f,g\right\rangle \triangleq \int
f(x)\,g(x)dx, $ and the multi-object exponential notation $h^{X}\triangleq %
\textstyle\prod\nolimits_{x\in X}h(x)$, where $h$ is a real-valued function,
with $h^{\varnothing }=1$ by convention \cite{Mahler07}. The cardinality (or
number of elements) of a finite set $X$ is denoted by $|X|$. Given a set $S$%
, $1_{S}(\cdot )$ denotes the indicator function of $S$, $\mathcal{F}(S)$
the class of finite subsets of $S$, and $S^{i}$ the $i^{\text{th}}$-fold
Cartesian product of $S$ with the convention $S^{0}=\{\varnothing \}$. We
introduce a generalization of the Kronecker delta that takes arguments such
as sets, vectors, etc., i.e.
\begin{equation*}
\delta _{Y}(X)\triangleq \left\{
\begin{array}{l}
1,\text{ if }X=Y \\
0,\text{ otherwise}%
\end{array}%
\right. ,
\end{equation*}%
A Gaussian \textit{Probability Density Function} (PDF) with mean $\mu $ and
covariance $\Sigma $ is denoted by $\mathcal{N}(\cdot ;\mu ,\Sigma )$.
Vectors are represented by lowercase letters, e.g. $x$, $\mathbf{x}$, while
finite sets are represented by uppercase letters, e.g. $X$, $\mathbf{X}$;
spaces are represented by blackboard bold letters e.g. $\mathbb{X}$, $%
\mathbb{Z}$, $\mathbb{L}$.

Given PDFs $p,$ $q$ and a scalar $\alpha >0$, the information fusion $\oplus
$ and weighting operators $\odot $ \cite{ccphd, auto2014, cpcl} are defined
as:
\begin{IEEEeqnarray}{rCl}
	\left( p\oplus q\right) (x) &\triangleq &\dfrac{p(x)\,q(x)}{\left\langle p,q\right\rangle }\,,  \label{oplus} \\
	\left( \alpha \odot p\right) (x) &\triangleq &\dfrac{\left[ p(x)\right]^{\alpha }}{\left\langle p^{\alpha },1\right\rangle }\,.  \label{odot}
\end{IEEEeqnarray}
For any PDFs $p,q$, $h$, and positive scalars $\alpha ,\beta $, the fusion
and weighting operators satisfy:
\begin{equation*}
\begin{array}{crcl}
\mbox{\textsc{p.a}\quad} & (p\oplus q)\oplus h & = & p\oplus (q\oplus h) = p
\oplus q\oplus h \\
\mbox{\textsc{p.b}\quad} & p\oplus q & = & q\oplus p \\
\mbox{\textsc{p.c}\quad} & (\alpha \,\beta ) \odot p & = & \alpha \odot
(\beta\odot p) \\
\mbox{\textsc{p.d}\quad} & 1 \odot p & = & p \\
\ \mbox{\textsc{p.e}\quad} & \alpha \odot (p\oplus q) & = & (\alpha \odot
p)\oplus(\alpha \odot q) \\
\mbox{\textsc{p.f}\quad} & (\alpha +\beta )\odot p & = & (\alpha \odot
p)\oplus(\beta \odot q)%
\end{array}%
\end{equation*}

\subsection{Network model}

\label{ssec:net}

\begin{figure}[h!]
\centering
\includegraphics[width=\columnwidth]{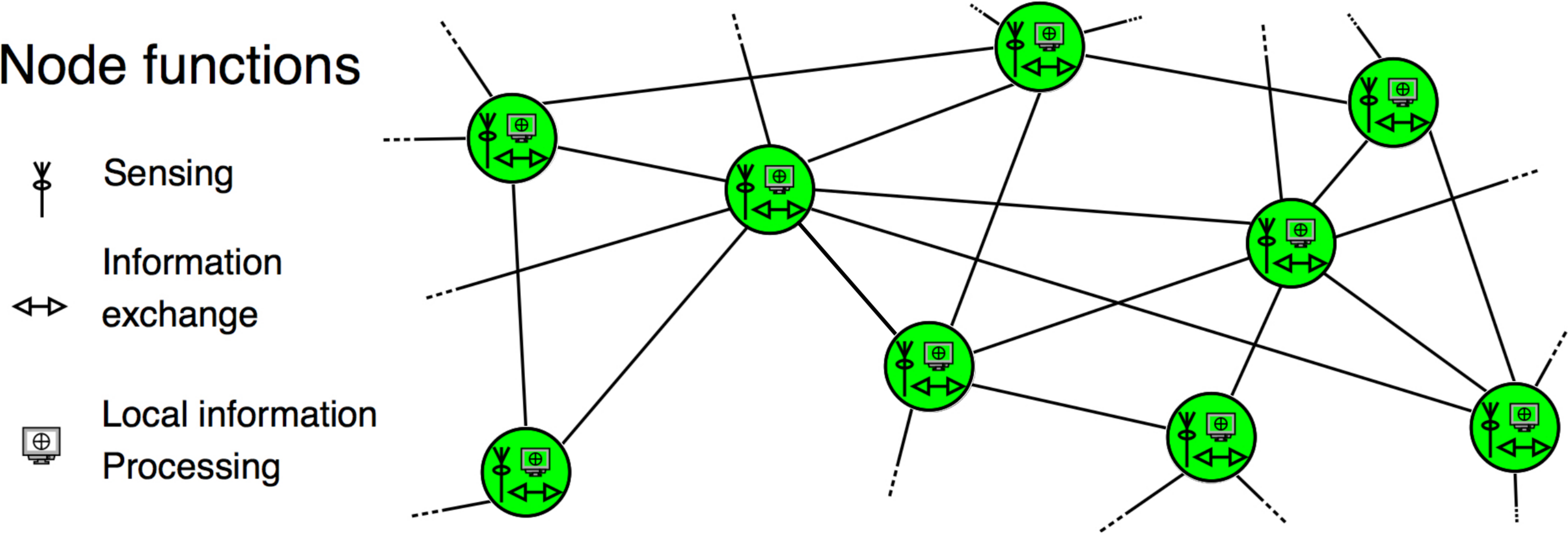}
\caption{Network model}
\label{net-model}
\end{figure}

The network considered in this work consists of heterogeneous and
geographically dispersed nodes having processing, communication and sensing
capabilities as depicted in Fig. \ref{net-model}. From a mathematical
viewpoint, the network is described by a directed graph $\mathcal{G}=\left(
\mathcal{N},\mathcal{A}\right) $ where $\mathcal{N}$ is the set of nodes and
$\mathcal{A}\subseteq \mathcal{N}\times \mathcal{N}$ the set of arcs,
representing \textit{links} (or \textit{connections}). In particular, $%
(i,j)\in \mathcal{A}$ if node $j$ can receive data from node $i$. For each
node $j\in \mathcal{N}$, $\mathcal{N}^{(j)}\triangleq \left\{ i\in \mathcal{N%
}:(i,j)\in \mathcal{A}\right\} $ denotes the set of in-neighbours (including
$j$ itself), i.e. the set of nodes from which node $j$ can receive data.

Each node performs local computations, exchanges data with the neighbors and
gathers measurements (e.g., angles, distances, Doppler shifts, etc.) of
objects present in the \textit{surrounding environment} (or \textit{%
surveillance area}). The network of interest has no \textit{central fusion
node} and its agents {operate without knowledge of the} network topology.

We are interested in networked estimation algorithms that are scalable with
respect to network size, and permit each node to operate without knowledge
of the dependence between its own information and the information from other
nodes.

\subsection{Distributed Single-Object Filtering and Fusion}

\label{ssec:dsof}

For single-object filtering, the problem of propagating information
throughout a sensor network $\left( \mathcal{N},\mathcal{A}\right) $ with neither
central fusion node nor knowledge of the network topology can be formalized
as follows.

The system model is described by the following Markov transition density and
measurement likelihood functions
\begin{gather}
f_{k|k-1}(x_{k}|x_{k-1})  \label{sys1} \\
g_{k}^{(i)}(z_{k}^{(i)}|x_{k})\,,i\in \mathcal{N}  \label{meas}
\end{gather}
The measurement at time $k$ is a vector $z_{k}=\left( z_{k}^{(1)}, \dots,
z_{k}^{(|\mathcal{N}|)}\right) $ of measurements from all $|\mathcal{N}|$
sensors, which are assumed to be conditionally independent given the state.
Hence the likelihood function of the measurement $z_{k}$ is given by
\begin{equation}
g_{k}(z_{k}|x_{k})=\prod_{i\in \mathcal{N}}g_{k}^{(i)}(z_{k}^{(i)}|x_{k})\,.
\end{equation}

Let $p_{k|k-1}(\cdot |z_{1:k-1})$ denote the prediction density of the state
at time $k$ given $z_{1:k-1}\triangleq \left( z_{1},\dots ,z_{k-1}\right) $,
and similarly $p_{k}(\cdot |z_{1:k})$ the posterior density of the state at
time $k$ given $z_{1:k}\triangleq \left( z_{1},\dots ,z_{k}\right) $. For
simplicity we omit the dependence on the measurements and write the
prediction and posterior densities respectively as $p_{k|k-1}$ and $p_{k}$.

In a \textit{centralized} setting, i.e. when the central node has access to
all measurements, the solution of the state estimation problem is given by
the Bayesian filtering recursion starting from a suitable initial prior $%
p_{0}$:
\begin{IEEEeqnarray}{rCl}
	p_{k|k-1}(x_{k}) &=&\left\langle f_{k|k-1}\!\left( x_{k}|\,\cdot \,\right), p_{k-1}\right\rangle \,,  \label{eq:STBayesPred} \\
	p_{k}(x_{k}) &=&\left( g_{k}(z_{k}|\,\cdot \,)\oplus p_{k|k-1}\right)(x_{k})\,.  \label{eq:STBayesUpdate}
\end{IEEEeqnarray}

On the other hand, in a \textit{distributed} setting each agent $i\in
\mathcal{N}$ updates its own posterior density $p_{k}^{(i)}$ by
appropriately fusing the available information provided by the subnetwork $%
\mathcal{N}^{(i)}$ (including node $i$). Thus, central to networked
estimation is the capability to fuse the posterior densities provided by
different nodes in a mathematically consistent manner. In this respect, the
information-theoretic notion of \textit{Kullback-Leibler Average} (KLA)
provides a consistent way of fusing PDFs \cite{auto2014}.

Given the PDFs $p^{(i)}$, $i\in \mathcal{I}$, and \textit{normalized
non-negative weights} $\omega ^{(i)}$ (i.e. non-negative
weights that sum up to $1$), $i\in \mathcal{I}$, the weighted\textit{\ Kullback-Leibler Average}
(KLA) $\overline{p}$ is defined as
\begin{equation}
\overline{p}=\arg \min_{p}~\displaystyle{\sum_{i\in \mathcal{I}}}~\omega
^{(i)}D_{KL}\left( p\parallel p^{(i)}\right)  \label{KLA}
\end{equation}%
where
\begin{equation}
D_{KL}\left( p\parallel p^{(i)}\right) =\int p\left( x\right) \log \!\left(
\dfrac{p(x)}{p^{(i)}(x)}\right) dx
\end{equation}%
is the \textit{Kullback-Leibler Divergence} (KLD) of $p^{(i)}$ from $p$. In
\cite{auto2014} it is shown that the weighted KLA in (\ref{KLA}) is the
\textit{normalized weighted geometric mean} of the PDFs, i.e.
\begin{equation}
\overline{p}(x)=\dfrac{\displaystyle\prod_{i\in \mathcal{I}}\left[ p^{(i)}(x)%
\right] ^{\omega ^{(i)}}}{\displaystyle\int \!\prod_{i\in \mathcal{I}}\!\!%
\left[ p^{(i)}(x)\right] ^{\omega ^{(i)}}\!\!\!\!dx}\,\triangleq \,%
\displaystyle{\bigoplus_{i\in \mathcal{I}}}\,\left( \omega ^{(i)}\odot
p^{(i)}\right) \!(x)\,.  \label{geo-mean}
\end{equation}%
Indeed, (\ref{geo-mean}) defines the well-known Chernoff fusion rule \cite%
{info,mori1}. Note that in the \textit{unweighted KLA} $\omega ^{(i)}=1/|%
\mathcal{I}|$, i.e.
\begin{equation}
\overline{p}=\displaystyle{\bigoplus_{i\in \mathcal{I}}}\,\dfrac{1}{|%
\mathcal{I}|}\odot p^{(i)}.  \label{eq:uwgeomean}
\end{equation}

\begin{rem}
The weighted KLA of Gaussians is also Gaussian \cite{auto2014}. More
precisely, let $(\Phi ,q)\triangleq (\Sigma ^{-1},\Sigma ^{-1}\mu )$ denote
the \textit{information (matrix-vector) pair }associated with $\mathcal{N}%
\left( \cdot ;\mu ,\Sigma \right) $, then the \textit{information pair} $(%
\overline{\Phi },\overline{q})$ of the KLA$\ \overline{p}(\cdot )=\mathcal{N}%
\left( \cdot ;\overline{\mu },\overline{\Sigma }\right) $ is the \textit{%
weighted arithmetic mean} of the information pairs $(\Phi ^{(i)},q^{(i)})$
of $p^{(i)}(\cdot )=\mathcal{N}\left( \cdot ;\mu ^{(i)},\Sigma ^{(i)}\right)
$. This is indeed the well-known \textit{Covariance Intersection} fusion
rule \cite{ci}.
\end{rem}

Having reviewed the fusion of PDFs via KLA, we next outline distributed
computation of the KLA via consensus.

\subsection{Consensus on PDFs}

\label{ssec:consensus}

{The idea behind consensus is to reach a \textit{collective agreement }(over
the entire network), by allowing each node }$i\in \mathcal{N}${\ to
iteratively update and pass its local information to neighbouring nodes \cite%
{Olfati}. Such repeated local operations provide a mechanism for propagating
information throughout the whole network. }In the context of this paper,
consensus is used (at each time step $k$) to perform distributed computation
of the collective unweighted KLA of the posterior densities $p_{k}^{(i)}$
over all nodes $i\in \mathcal{N}$.

Given the consensus weights $\omega ^{(i,j)}\in \lbrack 0,1]$ relating agent
$i$ to its in-neighbour nodes $j\in \mathcal{N}^{(i)}$, satisfying $%
\sum_{j\in \mathcal{N}^{(i)}}~\omega ^{(i,j)}=1$, suppose that, at time $k$,
each agent $i$ starts with the posterior $p_{k}^{(i)}$ as the initial
iterate $p_{k,0}^{(i)}$, and computes the $n^{th}$ consensus iterate by
\begin{equation}
p_{k,n}^{(i)}=\displaystyle{\bigoplus_{j\in \mathcal{N}^{(i)}}}\,\left(
\omega ^{(i,j)}\odot p_{k,n-1}^{(j)}\right)  \label{eq:consensus}
\end{equation}%
Then, using the properties of the operators $\oplus $ and $\odot $, it can
be shown that \cite{auto2014}
\begin{equation}
p_{k,n}^{(i)}=\displaystyle{\bigoplus_{j\in \mathcal{N}}}\,\left( \omega
_{n}^{(i,j)}\odot p_{k}^{(j)}\right)  \label{cons2}
\end{equation}%
where $\omega _{n}^{(i,j)}$ is the $(i,j)$-th entry of the square matrix $%
\Omega ^{n}$, and $\Omega $ is the consensus matrix with $(i,j)$-th entry
given by $\omega ^{(i,j)}1_{\mathcal{N}^{(i)}}(j)$ (it is understood that $%
p_{k}^{(j)} $ is omitted from the fusion whenever $\omega _{n}^{(i,j)}=0$).
Notice that (\ref{cons2}) expresses the local PDF in each node $i$ at consensus iteration $n$ as a weighted geometric mean of the initial local PDFs of all nodes.
More importantly, it was shown in \cite{Olfati,Xiao} that if the consensus
matrix $\Omega $ is primitive (i.e. with all non-negative entries and such
that there exists an integer $m$ such that $\Omega^m$ has all positive
entries) and doubly stochastic (all rows and columns sum up to 1), then for
any $i,j\in \mathcal{N}$
\begin{equation}
\lim_{n\rightarrow \infty }\omega _{n}^{(i,j)}=\dfrac{1}{|\mathcal{N}|}.
\end{equation}%
In other words, if the consensus matrix is primitive and doubly stochastic,
then the consensus iterate of each node approaches the collective unweighted
KLA of the posterior densities over the entire network as the number of
consensus steps tends to infinity \cite{Calafiore,auto2014}.

A necessary condition for $\Omega $ to be primitive \cite{Calafiore} is that
the associated network $\mathcal{G}$ be strongly connected, i.e. for any
pair of nodes $i$, $j\in \mathcal{N}$ there exists a directed path from $i$
to $j$ and vice versa. This condition is also sufficient when $\omega
^{(i,j)}>0$ for all $i \in \mathcal{N}$ and $j\in \mathcal{N}^{(i)}$. Further,
when $\mathcal{G}$ is undirected (i.e. whenever node $i$ receives
information from node $j$, it also sends information to $j$), choosing the
\textit{Metropolis weights}
\begin{equation}
\omega ^{(i,j)}\!=\!\left\{ \!\!\begin{array}{ll}
	\dfrac{1}{1\!+\!\operatorname{max}\!\left\{ |\mathcal{N}^{(i)}|,|\mathcal{N}^{(j)}|\right\} }, & \!\!i\in \mathcal{N},\,j\in \mathcal{N}^{(i)}\backslash\!\{i\} \\
	1-\sum_{j\in \mathcal{N}^{(i)}\backslash \{i\}}\omega ^{(i,j)}, & \!\!i\in\mathcal{N},\,j=i
	\end{array} \right.
\end{equation}
ensures that $\Omega $ is also doubly stochastic \cite{Xiao,Calafiore}.

In most tracking applications, the number of objects is
unknown and varies with time, while measurements are subjected to
misdetection, false alarms and association uncertainty.
This more general setting can be conveniently addressed by a rigorous
mathematical framework for dealing with multiple objects. Such a framework
is reviewed next, followed by the extension of the consensus methodology to
the multi-object realm.

\subsection{Labeled Random Finite Sets}

The RFS formulation of MOT provides the notion of \textit{multi-object
probability density }(for an unknown number of objects) \cite{VSD05} that
conceptually allows direct extension of the consensus methodology to
multi-object systems. Such a notion of multi-object probability density is
not available in the MHT or JPDA approaches \cite{reid, farina1,
farina2,barshalom2,blackmanpopoli}.

From a Bayesian estimation viewpoint the multi-object state is naturally
represented as a finite set, and subsequently modeled as an RFS \cite{VVPS10}%
. In this paper, unless otherwise stated we use the \textit{Finite Set STatistics}
(FISST) notion of integration/density to characterize RFSs \cite{Mahler07}.
While not a probability density \cite{Mahler07}, the FISST density is
equivalent to a probability density relative to an unnormalized distribution
of a Poisson RFS \cite{VSD05}.

Let $\mathbb{L}$ be a discrete space, and $\mathcal{L}:\mathbb{X}\mathcal{%
\times }\mathbb{L}\rightarrow \mathbb{L}$ be the projection defined by $%
\mathcal{L}((x,\ell ))=\ell $. Then $\mathcal{L}(\mathbf{x})$ is called the
label of the point $\mathbf{x}\in \mathbb{X}\mathcal{\times }\mathbb{L}$,
and a finite subset $\mathbf{X}$ of $\mathbb{X}\mathcal{\times }\mathbb{L}$
is said to have \emph{distinct labels} if and only if $\mathbf{X}$ and its
labels $\mathcal{L}(\mathbf{X})=\{\mathcal{L}(\mathbf{x}):\mathbf{x}\in
\mathbf{X}\}$ have the same cardinality. We define the \emph{distinct label
indicator }of $\mathbf{X}$ as $\Delta (\mathbf{X})\triangleq \delta _{|%
\mathbf{X}|}(|\mathcal{L(}\mathbf{X})|)$.

A \emph{labeled RFS} is an RFS over \ $\mathbb{X}\mathcal{\times }\mathbb{L}$
such that each realization has distinct labels. These distinct labels
provide the means to identify trajectories or tracks of individual objects
since a track is a time-sequence of states with the same label \cite{vovo1}.
The distinct label property ensures that at any time no two points can
share the same label, and hence no two trajectories can share any common
point in the extended space $\mathbb{X}\mathcal{\times }\mathbb{L}$. Hereinafter, symbols for labeled states and their distributions are
bolded to distinguish them from unlabeled ones, e.g. $\mathbf{x}$, $\mathbf{X%
}$, $\boldsymbol{\pi }$.

\subsubsection{Generalized Labeled Multi-Bernoulli (GLMB)}

A \textit{GLMB} \cite{vovo1} is a labeled RFS with state space $\mathbb{X}$
and (discrete) label space $\mathbb{L}$ distributed according to
\begin{equation}
\boldsymbol{\pi }(\mathbf{X})=\Delta (\mathbf{X})\sum_{\xi \in \Xi }w^{(\xi
)}(\mathcal{L}(\mathbf{X}))\left[ p^{(\xi )}\right] ^{\mathbf{X}}
\label{eq:glmb}
\end{equation}%
where $\Xi $ is a given discrete index set, each $p^{(\xi )}(\cdot ,\ell )$
is a PDF on $\mathbb{X}$, and each $w^{(\xi )}(L)$ is non-negative with
\begin{equation}
\sum_{\xi \in \Xi }\sum_{L\in \mathcal{F}\!\left( \mathbb{L}\right)
}w^{\left( \xi \right) }\!\left( L\right) =1\,.  \label{eq:vinc:1}
\end{equation}%
Each term in the mixture (\ref{eq:glmb}) consists of: a weight $w^{\left(
\xi \right) }\!\left( \mathcal{L}\!\left( \mathbf{X}\right) \right) $ that
only depends on the labels $\mathcal{L}\!\left( \mathbf{X}\right) $ of the
multi-object state $\mathbf{X}$; a multi-object exponential $\left[
p^{(\xi )}\right] ^{\mathbf{X}}$ that depends on the entire multi-object
state.

The cardinality distribution and intensity function (which is also the first
moment) of a GLMB are respectively given by%
\begin{align}
\Pr (\left\vert X\right\vert \text{=}n)& =\sum_{\xi \in \Xi }\sum_{I\in
\mathcal{F}\!\left( \mathbb{L}\right) }\delta _{n}(\left\vert I\right\vert )
\, w^{(\xi )}(I),  \label{eq:GLMBCard} \\
v(x,\ell )& =\sum_{\xi \in \Xi }p^{(\xi )}(x,\ell )\sum_{I\in \mathcal{F}%
\!\left( \mathbb{L}\right) }1_{I}(\ell ) \, w^{(\xi )}(I).
\end{align}

The GLMB is often written in the so-called $\delta $-GLMB form by using the
identity $w^{(\xi )}(J)=\sum_{I\in \mathcal{F}\!(\mathbb{L})}w^{(\xi
)}(I) \, \delta _{I}(J)$
\begin{equation}
\boldsymbol{\pi }(\mathbf{X})=\Delta (\mathbf{X})\sum_{\left( I,\xi \right)
\in \mathcal{F}\!\left( \mathbb{L}\right) \times \Xi } w^{\left( \xi
\right)} (I) \, \delta _{I}\!\left( \mathcal{L}\!\left( \mathbf{X}\right)
\right) \left[ p^{(\xi )}\right] ^{\mathbf{X}}  \label{eq:dglmb}
\end{equation}

For the standard multi-object system model that accounts for thinning,
Markov shifts and superposition, the GLMB family is a conjugate prior, and
is also closed under the Chapman-Kolmogorov equation \cite{vovo1}. Moreover,
the GLMB posterior can be tractably computed to any desired accuracy in the
sense that, given any $\epsilon >0$, an approximate GLMB within $\epsilon$ from the actual GLMB in  $L_{1}$ distance, can be computed (in
polynomial time) \cite{vovo2}.

\subsubsection{Marginalized $\protect\delta $-GLMB (M$\protect\delta $-GLMB)}

An M$\delta $-GLMB \cite{mdglmbf} is a special case of a GLMB with $\Xi =%
\mathcal{F}\!\left( \mathbb{L}\right) $ and density:
\begin{IEEEeqnarray}{rCl}
	\boldsymbol{\pi }(\mathbf{X}) & = & \Delta (\mathbf{X})\sum_{I\in \mathcal{F}\!\left( \mathbb{L}\right) }\delta _{I}(\mathcal{L}(\mathbf{X}))\,w\!\left(I\right)\left[ p\!\left( \cdot; I \right)\right] ^{\mathbf{X}}  \label{eq:mdglmb} \\
	& = & \Delta (\mathbf{X})w\!\left( \mathcal{L}(\mathbf{X})\right)\left[ p\!\left( \cdot; \mathcal{L}(\mathbf{X})\right) \right] ^{\mathbf{X}}.  \label{eq:mdglmb2}
\end{IEEEeqnarray}
An M$\delta $-GLMB is completely characterized by the parameter set $%
\{(w\!\left( I \right), p\!\left( \cdot; I \right) )$ : $I\in \mathcal{F}%
\!\left( \mathbb{L}\right) \}$, and for compactness we use the abbreviation $%
\boldsymbol{\pi }=\{(w\!\left( I \right), p\!\left( \cdot; I \right)
)\}_{I\in \mathcal{F}\!\left( \mathbb{L}\right) }$ for its density.

In \cite{mdglmbf}, an M$\delta $-GLMB of the form (\ref{eq:mdglmb}) was
proposed to approximate a $\delta $-GLMB of the form (\ref{eq:dglmb}), by
marginalizing (summing) over the discrete space $\Xi $, i.e. setting%
\begin{IEEEeqnarray}{rCl}
	w\!\left( I \right) & = &\sum_{\xi \in \Xi }w^{(\xi )}(I)\,,  \label{eq:mdglmb_w} \\
	p\!\left( x, \ell; I \right) & = &\frac{1_{I}(\ell )}{w\!\left( I \right)}\sum_{\xi \in \Xi }w^{(\xi)}(I) \, p^{(\xi )}(x,\ell )\,.  \label{eq:mdglmb_p}
\end{IEEEeqnarray}
Moreover, using a general result from \cite{papi2014} it was shown that such
M$\delta $-GLMB approximation mimimizes the KLD from the $\delta $-GLMB
while preserving the first moment and cardinality distribution \cite{mdglmbf}%
. The M$\delta $-GLMB approximation was used to develop a multi-sensor MOT
filter that is scalable with the number of sensors \cite{mdglmbf}.

\subsubsection{Labeled Multi-Bernoulli (LMB)}

An LMB \cite{vovo1} is another special case of a GLMB with density
\begin{equation}
\boldsymbol{\pi }(\mathbf{X})=\Delta (\mathbf{X})\left[ 1-r\right]^{\mathbb{M%
}\backslash \mathcal{L(}\mathbf{X})}\left[ 1_{\mathbb{M}}\,r\right] ^{%
\mathcal{L(}\mathbf{X})}p^{\mathbf{X}}  \label{eq:LMB2Convert}
\end{equation}%
An LMB is completely characterized by the (finite) parameter set $%
\{(r\!\left( \ell \right), p\!\left( \cdot, \ell \right))$ : $\ell \in
\mathbb{M}\}$, where $\mathbb{M}\subseteq \mathbb{L}$, $r\!\left( \ell
\right) \in \left[ 0,1\right] $ is the \textit{existence probability} of the
object with label $\ell $, and $p(\cdot ,\ell )$ is the PDF (on $\mathbb{X}$%
) of the object's state. For convenience we use the abbreviation $%
\boldsymbol{\pi }=\left\{ (r\!\left( \ell \right), p\!\left( \cdot, \ell
\right)) \right\} _{\ell \in \mathbb{M}}$ for the density of an LMB. In \cite%
{lmbf}, an approximation of a $\delta $-GLMB by an LMB with matching
unlabeled first moment was proposed together with an efficient MOT filter
known as the LMB filter.

\section{Information Fusion with labeled RFS}

\label{sec:lmbnwgm}

In this section, it is shown that the M$\delta $-GLMB and LMB densities are
algebraically closed under KL averaging, i.e. the KLAs of M$\delta $-GLMBs
and LMBs are respectively M$\delta $-GLMB and LMB. In particular we derive
closed form expressions for KLAs of M$\delta $-GLMBs and LMBs, which are
then used to develop consensus fusion of M$\delta $-GLMB and LMB posterior
densities.

\subsection{Multi-Object KLA}

\label{ssec:kla} The concept of probability density for the multi-object
state allows direct extension of the KLA notion to multi-object systems \cite%
{ccphd}.

Given the labeled multi-object densities $\boldsymbol{\pi }^{(i)}$ on $%
\mathcal{F}(\mathbb{X}\mathcal{\times }\mathbb{L})$, $i\in \mathcal{I}$, and
the normalized non-negative weights $\omega ^{(i)}$, $i\in \mathcal{I}$
(i.e. non-negative weights that sum up to $1$):

\subsubsection{The weighted\textit{\ KLA}}

$\overline{\boldsymbol{\pi }}$ is defined by
\begin{equation}
\overline{\boldsymbol{\pi }}\triangleq \arg \min_{\boldsymbol{\pi }%
}\sum_{i\in \mathcal{I}}\omega ^{(i)}D_{KL}\left( \boldsymbol{\pi }\parallel
\boldsymbol{\pi }^{(i)}\right)   \label{eq:kla}
\end{equation}%
where%
\begin{equation}
D_{\!KL\!}\left( \boldsymbol{\pi }\parallel \boldsymbol{\pi }^{(i)}\right)
\triangleq \int \boldsymbol{\pi }(\mathbf{X})\log \!\left( \dfrac{%
\boldsymbol{\pi }(\mathbf{X})}{\boldsymbol{\pi }^{(i)}(\mathbf{X})}\right)
\delta \mathbf{X}  \label{eq:dkl}
\end{equation}%
is the KLD of $\boldsymbol{\pi }^{(i)}$ from $\boldsymbol{\pi}$ \cite%
{MahlerPHD2, Mahler07}, and the integral is the FISST \textit{set integral}
defined for any function $f\!$ on $\!\mathcal{F}(\mathbb{X}\mathcal{\times }%
\mathbb{L})$ by
\begin{IEEEeqnarray}{l}
	\int \!f\!(\mathbf{X})\delta \mathbf{X} = \sum_{i=0}^{\infty }\frac{1}{i!}\!\!\sum_{\left( \ell _{1},\dots ,\ell_{i}\right) \in \mathbb{L}^{i}}\int \!\!f\!(\{\left( x_{1},\ell _{1}\right),\dots ,\left( x_{i},\ell _{i}\right) \})\,d(x_{1},...,x_{i}).  \label{eq:setint}
\end{IEEEeqnarray}
Note that the integrand $f(\mathbf{X})$ has unit of $K^{-|\mathbf{X}|}$,
where $K$ is the unit of hyper-volume on $\mathbb{X}$. For compactness, the
inner product notation $\left\langle f,g\right\rangle $ will be used also for the
set integral $\int \!\!f(\mathbf{X})g(\mathbf{X})\delta \mathbf{X}$, when $g(%
\mathbf{X})$ has unit independent of $|\mathbf{X}|$.

\subsubsection{The normalized weighted geometric mean}

is defined by%
\begin{equation}
\bigoplus_{i\in \mathcal{I}}\left( \omega ^{(i)}\odot \boldsymbol{\pi }%
^{(i)}\right) =\dfrac{\displaystyle\prod_{i\in \mathcal{I}}\left[
\boldsymbol{\pi }^{(i)}(\mathbf{X})\right] ^{\!\omega ^{(i)}}}{\displaystyle%
\int \!\prod_{i\in \mathcal{I}}\left[ \boldsymbol{\pi }^{(i)}(\mathbf{X})%
\right] ^{\!\omega ^{(i)}}\!\!\!\delta \mathbf{X}}\,,  \label{eq:monwgm}
\end{equation}%
Note that since the exponents $\omega ^{(i)}$, $i\in \mathcal{I}$, sum up to
unity, the product in the numerator of (\ref{eq:monwgm}) has unit of $K^{-|%
\mathbf{X}|}$, and the set integral in the denominator of (\ref{eq:monwgm})
is well-defined and unitless. Hence, the normalized weighted geometric mean (%
\ref{eq:monwgm}), originally proposed by Mahler in \cite{gci} as the
multi-object Chernoff fusion rule, is well-defined.

Similar to the single object case, the weighted KLA is given by the
normalized weighted geometric mean.

\begin{thm}
\label{thm:kla} \cite{ccphd} - Given multi-object densities $\boldsymbol{\pi
}^{(i)}$, $i\in \mathcal{I}$, and normalized non-negative weights $\omega
^{(i)}$, $i\in \mathcal{I}$,
\begin{equation}
\!\!\arg \min_{\boldsymbol{\pi }}\sum_{i\in \mathcal{I}}\!\omega ^{(i)\!} \,
D_{\!KL}\!\left( \boldsymbol{\pi }\!\parallel \boldsymbol{\pi }%
^{(i)\!}\right) =\bigoplus_{i\in \mathcal{I}}\!\left( \omega ^{(i)}\odot
\boldsymbol{\pi }^{(i)\!}\right) \!.  \label{eq:kla:gci}
\end{equation}
\end{thm}

Note that the label space $\mathbb L$ has to be the same for all the densities $\boldsymbol{\pi}^{(i)}$ for the KLA to be well-defined. In \cite[Theorem 5.1]{SPIE}, it has been mathematically proven that, due to the weight normalization $\sum_i \omega^{(i)} =1$, the weighted geometric mean (\ref{eq:kla:gci})
ensures immunity to the double counting of information irrespective of the unknown common information in the densities $\boldsymbol{\pi}^{(i)}$.

In \cite{ccphd}, it was shown that Poisson and \textit{independently
identically distributed cluster} (IID-cluster) RFSs are algebraically closed
under KL averaging. While the GLMB family is algebraically closed under the
Bayes recursion for the standard multi-object system model and enjoys a
number of useful analytical properties, it is not algebraically closed under
KL averaging. Nonetheless, there are versatile subfamilies of the GLMBs that
are algebraically closed under KL averaging.

\subsection{Weighted KLA of M$\protect\delta $-GLMB Densities}

The following result states that the KLA of M$\delta $-GLMB densities is
also an M$\delta $-GLMB density. The proof is provided in Appendix A.

\begin{pro}
\label{pro:mdglmb:fusion}Given M$\delta$-GLMB densities
$\boldsymbol{\pi}^{(i)}=\{(w^{(i)}(I),p^{(i)}(\cdot; I))\}_{I\in \mathcal{F}%
\!\left( \mathbb{L}\right) }$, $i\in \mathcal{I}$, and normalized
non-negative weights $\omega^{(i)}$, $i\in \mathcal{I}$, the normalized
weighted geometric mean $\overline{\boldsymbol{\pi}}$, and hence the KLA, is
an M$\delta $-GLMB given by:
\begin{equation}
\overline{\boldsymbol{\pi }} =\left\{ (\overline{w}\!\left( L \right) ,%
\overline{p}\!\left( \cdot; L\right) )\right\} _{L\in \mathcal{F}\!\left(
\mathbb{L}\right) }  \label{eq:mdglmb:nwgm}
\end{equation}
where
\begin{IEEEeqnarray}{rCl}
	\overline{w}\!\left( L \right) & = & \frac{\displaystyle\prod_{i\in \mathcal{I}}\!\!\left( w^{\left( i \right)\!}\left( L \right)\right)^{\!\omega^{(i)}}\!\!\left[ \int \!\!\prod_{i\in \mathcal{I}}\!\!\left( p^{\left( i \right) \!}\!\left( x,\cdot; L \right) \!\right)^{\omega^{(i)}}\!\!\!dx\!\right] ^{\!L}}{\displaystyle\sum_{J\subseteq\mathbb{L}}\prod_{i\in \mathcal{I}}\!\!\left( w^{\left( i\right)\!}\left( J \right)\right) ^{\!\omega ^{(i)}}\!\!\left[ \int \!\!\prod_{i\in \mathcal{I}}\!\!\left( p^{\left( i\right) \!}\!\left( x,\cdot; J \right) \!\right)^{\!\omega ^{(i)}}\!\!\!dx\!\right] ^{\!J}} \IEEEeqnarraynumspace\\
	\overline{p} \!\left( \cdot ,\ell; L\right) & = & \frac{\displaystyle\prod_{i\in \mathcal{I}}\!\!\left( p^{\left( i\right) \!}\left( \cdot ,\ell; L \right) \!\right)^{\!\omega ^{(i)}}}{\displaystyle\int \!\prod_{i\in \mathcal{I}}\!\!\left(p^{\left( i\right) \!}\!\left( x, \ell; L \right) \!\right) ^{\!\omega^{(i)}}\!\!\!dx} \, .
\end{IEEEeqnarray}
\end{pro}

\begin{rem}
The component
$(\overline{w}\!\left( L\right) ,\overline{p}\!\left( \cdot ;L\right) )$ of
the KLA M$\delta $-GLMB can be rewritten as
\begin{IEEEeqnarray}{rCl}
	\overline{w}\!\left( L\right) & \propto & \prod_{i\in\mathcal{I}}\!\!\left( w^{\left( i\right) \!}\left( L \right)\right) ^{\!\omega^{(i)}}\!\!\left[ \int \!\!\prod_{i\in \mathcal{I}}\!\!\left( p^{\left( i \right) }\!\left( x,\cdot; L \right) \!\right) ^{\!\omega ^{(i)}}\!\!\!dx\!\right] ^{\!L}   \label{eq:mdglmbfusion:w} \IEEEeqnarraynumspace\\
	\overline{p}(\cdot; L ) &=& \bigoplus_{i\in \mathcal{I}}\left( \omega ^{(i)}\odot p^{\left( i\right)}(\cdot; L )\right)
	\label{eq:mdglmbfusion:pdf}
\end{IEEEeqnarray}where (\ref{eq:mdglmbfusion:pdf}) is indeed the Chernoff
fusion rule for the single-object PDFs \cite{ci}. Note also from (\ref%
{eq:mdglmbfusion:w}) and (\ref{eq:mdglmbfusion:pdf}) that each M$\delta $-GLMB component $(\overline{w}\!\left( L\right) ,\overline{p}\!\left( \cdot ;L\right) )$ can
be independently determined. Thus, the overall fusion procedure is fully
parallelizable.
\end{rem}

\subsection{Weighted KLA of LMB Densities}

The following result states that the KLA of LMB densities is also an LMB
density. The proof is provided in Appendix B.

\begin{pro}
\label{pro:lmb:fusion}Given LMB densities
$\boldsymbol{\pi }^{(i)}=\{(r^{(i)}(\ell),p^{(i)}(\cdot, \ell))\}_{\ell \in
\mathbb{L}}$, $i\in\mathcal{I}$, and normalized non-negative weights $\omega
^{(i)}$, $i\in\mathcal{I}$, the normalized weighted geometric mean $%
\overline{\boldsymbol{\pi}}$, and hence the KLA, is an LMB given by:

\begin{equation}
\overline{\boldsymbol{\pi }} =\left\{ \left( \overline{r}\!\left( \ell
\right), \overline{p}\!\left( \cdot, \ell \right) \right) \right\} _{\ell
\in \mathbb{L}}  \label{eq:lmb:nwgm}
\end{equation}
where
\begin{IEEEeqnarray}{rCl}
	\overline{r}\!\left( \ell \right) & = & \dfrac{\displaystyle\int\!\!\prod\limits_{i\in \mathcal{I}}\!\!\left( r^{(i)}(\ell) p^{(i)}(x, \ell)\right)^{\!\omega^{(i)}}\!\!\!dx}{\displaystyle\prod\limits_{i\in \mathcal{I}}\!\!\left( 1 - r^{(i)}(\ell)\right)^{\!\omega^{(i)}}\!\!\!\!+\int\!\!\prod\limits_{i\in \mathcal{I}}\!\!\left( r^{(i)}(\ell) p^{(i)}(x, \ell)\right)^{\!\omega^{(i)}}\!\!\!\!dx} \label{eq:lmbfusion:ex} \IEEEeqnarraynumspace\\
	\overline{p}\!\left( \cdot, \ell \right) & = & \bigoplus_{i\in \mathcal{I}}\left( \omega ^{(i)}\odot p^{(i)}( \cdot, \ell )\right) \, .  \label{eq:lmbfusion:pdf}
\end{IEEEeqnarray}
\end{pro}

\begin{rem}
Similar to the KLA of M$\delta $-GLMBs, (\ref{eq:lmbfusion:pdf}) is indeed
the Chernoff fusion rule for the single-object PDFs \cite{ci}. Note from (\ref{eq:lmbfusion:ex})
and (\ref{eq:lmbfusion:pdf}) that each LMB component
$(\overline{r}\!\left( \ell \right) ,\overline{p}\left( \cdot ,\ell \right) )
$ can be independently determined. Thus, the overall fusion procedure is
fully parallelizable.
\end{rem}

\subsection{Consensus Fusion for Labeled RFSs}

\label{sec:fmttf}Consider a sensor network $\mathcal{N}$ with multi-object
density $\boldsymbol{\pi }^{(i)}$\ at each node $i$, and non-negative
consensus weights $\omega ^{(i,j)}$ relating node $i$ to nodes $j\in
\mathcal{N}^{(i)}$, such that $\sum_{j\in \mathcal{N}^{(i)}}~\omega
^{(i,j)}=1$. The global KLA over the entire network can be computed in a
distributed and scalable way by using the consensus algorithm \cite[Section
III.A]{ccphd,auto2014}. Starting with $\boldsymbol{\pi }_{0}^{(i)}=%
\boldsymbol{\pi }^{(i)}$, each node $i\in \mathcal{N}$ carries out the
consensus iteration
\begin{equation}
\boldsymbol{\pi }_{n}^{(i)}=\bigoplus_{j\in \mathcal{N}^{(i)}}\left( \omega
^{(i,j)}\odot \boldsymbol{\pi }_{n-1}^{(j)}\right) \,.
\label{eq:consenus:alg}
\end{equation}

As shown in \cite[Section III-B]{ccphd}, the consensus iteration (\ref{eq:consenus:alg})---which is the multi-object
counterpart of equation (\ref{eq:consensus})---enjoys some nice convergence properties.
In particular,
if the consensus matrix
is primitive and doubly stochastic, the consensus iterate of each node in
the network converges to the global unweighted KLA of the multi-object
posterior densities as $n$ tends to infinity. Convergence analysis for the
multi-object case follows along the same line as in \cite{Calafiore,auto2014} since $\mathcal{F}\!\left( \mathbb{X}\times \mathbb{L}%
\right) $ is a metric space \cite{Mahler07}. In practice, the iteration is
stopped at some finite $n$.
Further, as pointed out in \cite[Remark 1]{SPIE},
the consensus iterations (\ref{eq:consenus:alg}) always generate multi-object densities $\boldsymbol{\pi}_n^{(i)}$ that mitigate double counting irrespectively of the number
$n$ of iterations.

Starting with $\delta $-GLMBs, the consensus iteration (\ref{eq:consenus:alg}%
) always returns M$\delta $-GLMBs, moreover the M$\delta $-GLMB parameter
set can be computed by the \textit{M}$\delta $\textit{-GLMB fusion rules} (%
\ref{eq:mdglmbfusion:w}) and (\ref{eq:mdglmbfusion:pdf}). Similarly, for
LMBs the consensus iteration (\ref{eq:consenus:alg}) always returns LMBs
whose parameter set can be computed by the \textit{LMB fusion rules} (\ref%
{eq:lmbfusion:ex}) and (\ref{eq:lmbfusion:pdf}). The fusion rules (\ref%
{eq:mdglmbfusion:pdf}) and (\ref{eq:lmbfusion:pdf}) involve consensus of
single-object PDFs.

A typical choice for representing each single-object density is a \textit{%
Gaussian Mixture} (GM) \cite{phd,cphd}.
In this case, the fusion rules (\ref{eq:mdglmbfusion:pdf}) and (\ref%
{eq:lmbfusion:pdf}) involve exponentiation and multiplication of GMs where
the former, in general, does not provide a GM. Hence, in order to preserve
the GM form, a suitable approximation of the GM exponentiation has to be
devised. The in-depth discussion and efficient implementation proposed in
\cite[Section III.D]{ccphd} for generic GMs can also be applied to the
location PDF fusion (\ref{eq:mdglmbfusion:pdf}) and (\ref{eq:lmbfusion:pdf}%
). Considering, for the sake of simplicity, the case of two GMs
\begin{equation*}
p_i(x) = \sum_{j=1}^{N_i} \alpha_{i,j} \mathcal{N} \left( x;\mu_{i,j},
P_{i,j} \right)
\end{equation*}
for $i \in \{ a,b \}$, (\ref{eq:mdglmbfusion:pdf}) and (\ref%
{eq:lmbfusion:pdf}) can be approximated as follows:
\begin{equation}
\overline{p}(x) = \dfrac{\displaystyle{\sum_{j=1}^{N_{a}}} \displaystyle{%
\sum_{k=1}^{N_{b}}} \, \overline{\alpha}_{jk} \,\mathcal{N} \left( x;
\overline{\mu}_{jk}, \overline{P}_{jk} \right) }{\displaystyle{%
\sum_{j=1}^{N_{a}}}\displaystyle{\sum_{k=1}^{N_{b}}} \, \overline{\alpha}%
_{jk}}  \label{f1}
\end{equation}
where
\begin{IEEEeqnarray}{rCl}
	\overline{P}_{jk} &=&\left[ \omega P_{a,j}^{-1} + (1-\omega ) P_{b,k}^{-1} \right] ^{-1}  \label{f2} \\
	\overline{\mu}_{jk} &=& P_{jk} \left[ \omega P_{a,j}^{-1} \mu_{a,j} + (1-\omega ) P_{b,k}^{-1} \mu_{b,k} \right]\label{f3} \\
	\overline{\alpha}_{jk} &=& \alpha _{a,j}^{\omega } \, \alpha_{b,k}^{1-\omega } \beta \!\left( \omega, P_{a,j} \right) \beta\!\left( 1-\omega , P_{b,k} \right) \mathcal{N}\!\left( \mu_{a,j}-\mu_{b,k};0,\frac{P_{a,j}}{\omega} + \frac{P_{b,k}}{1-\omega } \right)   \label{f4} \\
	\beta \!\left( \omega, P \right)  &\triangleq &\dfrac{\left[ \operatorname{det}\!\left( 2\pi P\omega ^{-1}\right) \right] ^{\frac{1}{2}}}{\left[ \operatorname{det} \!\left( 2\pi P\right) \right] ^{\frac{\omega }{2}}}  \label{f5}
\end{IEEEeqnarray}
The fusion (\ref{f1}) can be extended to $\left\vert \mathcal{N}\right\vert
\geq 2$ agents by sequentially applying the pairwise fusion rule (\ref{f1})-(%
\ref{f5}) $\left\vert \mathcal{N}\right\vert -1$ times. By the associative
and commutative properties of multiplication, the ordering of pairwise
fusions is irrelevant. Notice that (\ref{f1})-(\ref{f5}) amounts to
performing a Chernoff fusion on any possible pair formed by a Gaussian
component of agent $a$ and a Gaussian component of agent $b$. Moreover, the
coefficient $\overline{\alpha}_{jk}$ of the resulting (fused) component
includes a factor $\mathcal{N}\left(
\mu_{a,j}-\mu_{b,k};0,\omega^{-1}P_{a,j}+(1-\omega )^{-1}P_{b,k}\right) $
that measures the separation of the two fusing components $\left(
\mu_{a,j},P_{a,j} \right)$ and $\left( \mu_{b,k},P_{b,k} \right)$.
The approximation (\ref{f1})-(\ref{f5}) is reasonably accurate for well-separated Gaussian components but might easily deteriorate in presence of closely located components.
In this respect, merging of nearby components before fusion has been exploited in \cite{ccphd} to mitigate the problem.
Further, a more accurate, but also more computationally demanding, approximation has been proposed in \cite{gunay}.

The other common approach for approximating a single object PDF $p(\cdot )$
is via \textit{particles}, i.e. weighted sums of Dirac delta functions,
which can address non-linear, non-Gaussian dynamics and measurements as well
as non-uniform field of view. However, computing the KLA requires
multiplying together powers of relevant PDFs, which cannot be performed
directly on weighted sums of Dirac delta functions. While this problem can
be addressed by further approximating the particle PDFs by continuous PDFs
(e.g. GMs) using techniques such as kernel density estimation \cite{emd},
least square estimation \cite{clike1,clike2} or parametric model approaches
\cite{coates}, such approximations increase the in-node computational
burden. Moreover, the local filtering steps are also more resource demanding
compared to a GM implementation. Hence, at this developmental stage, it is
more efficient to work with GM approximations.

\section{Consensus DMOT}

\label{sec:clmbif} In this section, we present two novel fully distributed
and scalable multi-object tracking algorithms based on Propositions \ref%
{pro:mdglmb:fusion} and \ref{pro:lmb:fusion} along with consensus \cite%
{Olfati,Xiao,Calafiore,auto2014} to propagate information throughout the
network.

\subsection{Bayesian Multi-Object Filtering}

\label{subsec:RFSBayes}

We begin this section with the Bayes MOT filter that propagates the
multi-object posterior/filtering density. In this formulation the
multi-object state is modeled as a labeled RFS in which a label is an
ordered pair of integers $\ell =(k,i)$, where $k$ is the \textit{time of
birth}, and $i\in \mathbb{N}$ is a unique index to distinguish objects born
at the same time. The label space for objects born at time $k$ is $\mathbb{L}%
_{k}=\{k\}\times \mathbb{N}$. An object born at time $k$ has, therefore,
state $\mathbf{x}\in \mathbb{X}\mathcal{\times }\mathbb{L}_{k}$. Hence, the
label space for objects at time $k$ (including those born prior to $k$),
denoted as $\mathbb{L}_{0:k}$, is constructed recursively by $\mathbb{L}%
_{0:k}=\mathbb{L}_{0:k-1}\cup \mathbb{L}_{k}$ (note that $\mathbb{L}_{0:k-1}$
and $\mathbb{L}_{k}$ are disjoint). A multi-object state $\mathbf{X}$ at
time $k$, is a finite subset of $\mathbb{X}\mathcal{\times }\mathbb{L}_{0:k}$%
. For convenience, we denote $\mathbb{L}_{-}\mathbb{\triangleq L}_{0:k-1}$, $%
\mathbb{B\triangleq L}_{k}$, and $\mathbb{L}\triangleq \mathbb{L}_{-}\cup
\mathbb{B}$.

Let $\boldsymbol{\pi }_{k}$ denote the \textit{multi-object filtering density%
} at time $k$, and $\boldsymbol{\pi }_{k|k-1}$ the \textit{multi-object
prediction density} (for compactness, the dependence on the measurements is
omitted). Then, starting from $\boldsymbol{\pi }_{0}$, the \textit{%
multi-object Bayes recursion} propagates $\boldsymbol{\pi }_{k}$ in time
according to the following update and prediction \cite{MahlerPHD2, Mahler07}
\begin{IEEEeqnarray}{rCl}
	\boldsymbol{\pi }_{k|k-1}(\mathbf{X}_{k}) &=&\left\langle \boldsymbol{f}_{k|k-1}(\mathbf{X}_{k}|\,\cdot \,),\boldsymbol{\pi }_{k-1}(\,\cdot \,)\right\rangle \, ,  \label{eq:MTBayesPred} \\
	\boldsymbol{\pi }_{k}(\mathbf{X}_{k}) &=&\left( g_{k}(Z_{k}|\,\cdot\,)\oplus \boldsymbol{\pi }_{k|k-1}(\,\cdot \,)\right) \left( \mathbf{X}_{k}\right) \, ,  \label{eq:MTBayesUpdate}
\end{IEEEeqnarray}
where $\boldsymbol{f}_{k|k-1}(\,\cdot \,|\,\cdot \,)$ is the \textit{%
multi-object transition density} from time $k-1$ to time $k$, and $%
g_{k}(\,\cdot \,|\,\cdot \,)$ is the \textit{multi-object likelihood}
\textit{function} at time $k$. The multi-object likelihood function
encapsulates the underlying models for detections and false alarms while the
multi-object transition density encapsulates the underlying models of
motion, birth and death. The multi-object filtering (or posterior) density
captures all information on the number of objects, and their states \cite%
{Mahler07}.

Note that the recursions (\ref{eq:MTBayesPred})-(\ref{eq:MTBayesUpdate}) are
the multi-object counterpart of (\ref{eq:STBayesPred})-(\ref%
{eq:STBayesUpdate}), which admit a closed form solution, under the standard
multi-object system model, known as the GLMB filter \cite{vovo1} (see also
\cite{vovo2} for implementation details). However, the GLMB family is not
closed under KL averaging. Consequently, we look towards approximations such
as the M$\delta $-GLMB and LMB filters for analytic solutions to DMOT.

\subsection{The M$\protect\delta$-GLMB Filter}

\label{ssec:mdglmbf} In the following we outline the prediction and update
steps for the M$\delta $-GLMB filter. Additional details can be found in
\cite{vovo2}.

\subsubsection{M$\protect\delta$-GLMB Prediction}

\label{ssec:mdglmb:pred} Given the previous multi-object state $\mathbf{X}%
_{k-1}$, each state $(x_{k-1},\ell _{k-1})\in \mathbf{X}_{k-1}$ either
continues to exist at the next time step with probability $%
P_{S}(x_{k-1},\ell _{k-1})$ and evolves to a new state $(x_{k},\ell _{k})$
with probability density $f_{k|k-1}(x_{k}|x_{k-1},\ell _{k-1})$, or dies
with probability $1-P_{S}(x_{k-1},\ell _{k-1})$. The set of new objects born
at the next time step is distributed according to the LMB
\begin{equation}
\boldsymbol{f}_{B}(\mathbf{X})=\Delta (\mathbf{X})\left[ 1-r_{B}\right] ^{%
\mathbb{B}\backslash \mathcal{L(}\mathbf{X})}\left[ 1_{\mathbb{B}} \, r_{B} %
\right]^{\mathcal{L(}\mathbf{X})}\left[ p_{B}\right] ^{\mathbf{X}}.
\label{eq:Birth_transition1}
\end{equation}
It is assumed that $p_{B}(\cdot ,l)\neq p_{B}(\cdot ,j)$ when $l\neq j$.
Note that $\boldsymbol{f}_{B}(\mathbf{X})=0$ if $\mathbf{X}$ contains any
element $\mathbf{x}$ with $\mathcal{L}\left( \mathbf{x}\right) \notin
\mathbb{B}$. The multi-object state at the next time $\mathbf{X}$ is the
superposition of surviving objects and new born objects, and the
multi-object transition density can be found in \cite[Subsection IV.D]{vovo1}%
.

\begin{rem}
\label{rem:label} The LMB birth model assigns unique labels to objects in
the following sense. Consider two objects born at time $k$ with kinematic
states $x$ and $y$. In birth models such as labeled Poisson \cite{vovo1}, $x$
could be assigned label $(k,1)$ and $y$ label $(k,2)$, i.e. the multi-object
state is $\{(x,(k,1)),(y,(k,2))\}$, or conversely $x$ assigned label $(k,2)$
and $y$ label $(k,1)$, i.e. the multi-object state is\ $%
\{(x,(k,2)),(y,(k,1))\}$. Such non-uniqueness arises because the kinematic
state of an object is generated independently of the label. This does not
occur in the LMB model because an object with label $\ell $, has kinematic
state generated from $p_{B}(\cdot ,\ell )$. If kinematic states $x$ and $y$
are drawn respectively from $p_{B}(\cdot ,(k,1))$ and $p_{B}(\cdot ,(k,2))$,
then the labeled multi-object state is uniquely $\{(x,(k,1)),(y,(k,2))\}$.
\end{rem}

Given the M$\delta$-GLMB multi-object posterior density $\boldsymbol{\pi }%
_{k-1}=\{(w_{k-1}\!\left( I \right), p_{k-1}\!\left( \cdot; I \right)
)\}_{I\in \mathcal{F}\!\left( \mathbb{L}_{-}\right)}$, the multi-object
prediction density is the M$\delta $-GLMB $\boldsymbol{\pi }%
_{k|k-1}=\{(w_{k|k-1}\!\left( I \right), p_{k|k-1}\!\left( \cdot; I \right)
)\}_{I\in\mathcal{F}\!\left( \mathbb{L}\right) }$, where
\begin{IEEEeqnarray}{rCl}
	w_{k|k-1}\!\left( I \right) & = & \left[1-r_{B}\right] ^{\mathbb{B}\backslash I}\!\left[ 1_{\mathbb{B}}\,r_{B}\right] ^{\!I\cap \mathbb{B}} w_{S}^{(I)\!}(I\cap {\mathbb{L}}_{-}) \\
	p_{k|k-1\!}\!\left(x,\ell; I \right) & = & 1_{\mathbb{B}}(\ell )p_{B}(x,\ell )+1_{\mathbb{L}_{-}}(\ell )p_{S}(x,\ell; I)  \\
	w_{S}^{(I)}(L) & = & \left[ \overline{P}_{S}^{(I)}\right]^{L}\sum_{J\supseteq L}\left[ 1-\overline{P}_{S}^{(I)}\right] ^{J-L}w_{k-1}\!\left( J \right) \IEEEeqnarraynumspace\\
	p_{S}(x,\ell; I ) & = & \frac{\left\langle P_{S}(\cdot ,\ell )f_{k|k-1}(x|\cdot ,\ell ),p_{k-1}\!\left( \cdot ,\ell, I \right)\right\rangle }{\overline{P}_{S}^{(I)}(\ell )}  \\
	\overline{P}_{S}^{(I)}(\ell ) & = & \left\langle P_{S}(\cdot ,\ell ),p_{k-1}\!\left( \cdot ,\ell; I \right) \right\rangle \, .
\end{IEEEeqnarray}

\subsubsection{M$\protect\delta$-GLMB Update}

\label{ssec:mdglmb:corr} Given a multi-object state $\mathbf{X}_{k}$, each
state $(x_{k},\ell _{k})\in \mathbf{X}_{k}$ is either detected with
probability $P_{D}\left( x_{k},\ell _{k}\right) $ and generates a
measurement $z$ with likelihood $g_{k}(z|x_{k},\ell _{k})$, or missed with
probability $1-P_{D}(x_{k},\ell _{k})$. The multi-object observation $%
Z_{k}=\{z_{1},\dots ,z_{|Z_{k}|}\}$ is the superposition of the detected
points and Poisson clutter with intensity function $\kappa $. Assuming that,
conditional on $\mathbf{X}_{k}$, detections are independent, and that
clutter is independent of the detections, the multi-object likelihood is
given by \cite[Subsection IV.D]{vovo1}
\begin{equation}
g_{k}(Z_{k}|\mathbf{X}_{k})\propto \sum_{\theta \in \Theta (\mathcal{L(}%
\mathbf{X}_{k}))}\left[ \psi _{Z_{k}}(\cdot ;\theta )\right] ^{\mathbf{X}%
_{k}}  \label{eq:RFSmeaslikelihood0}
\end{equation}%
where $\Theta (I)$ is the set of mappings $\theta :I\rightarrow \{0,1,\dots
,|Z_{k}|\},$ such that $\theta (i)=\theta (i^{\prime })>0$ implies $%
i=i^{\prime }$, and
\begin{equation*}
\psi _{Z_{k}}(x,\ell ;\theta )=\left\{
\begin{array}{ll}
\dfrac{P_{D}(x,\ell )\,g_{k}(z_{\theta (\ell )}|x,\ell )}{\kappa (z_{\theta
(\ell )})}\,, & \text{if }\theta (\ell )>0 \\
1-P_{D}(x,\ell )\,, & \text{if }\theta (\ell )=0%
\end{array}%
\right. \,.\ \
\end{equation*}%
Note that an association map $\theta $ specifies which tracks generated
which measurements, i.e. track $\ell $ generates measurement $z_{\theta
(\ell )}\in Z_{k}$, with undetected tracks assigned to $0$. The condition
\textquotedblleft $\theta (i)=\theta (i^{\prime })>0$ implies $i=i^{\prime }$%
\textquotedblright , means that, at any time, a track can generate at most
one measurement, and a measurement can be assigned to at most one track.

Given the M$\delta $-GLMB multi-object prediction density $\boldsymbol{\pi}%
_{k|k-1}=\{(w_{k|k-1}\!\left( I \right), p_{k|k-1}\!\left( \cdot; I \right)
)\}_{I\in \mathcal{F}\!\left( \mathbb{L}\right) }$, the M$\delta $-GLMB
updated density is given by $\boldsymbol{\pi }_{k}=\{(w_{k}\!\left( I
\right), p_{k}\!\left( \cdot; I \right) )\}_{I\in \mathcal{F}\!\left(
\mathbb{L}\right) }$, where
\begin{IEEEeqnarray}{rCl}
	w_{k}\!\left( I \right) & = &\sum_{\theta \in \Theta (I)}w_{k}^{\left( I, \theta \right)} \, , \label{eq:MGLMB_w} \\
	p_{k}\!\left( x, \ell; I \right) & = &\frac{1_{I}(\ell )}{w_{k}\!\left( I \right)} \sum_{\theta \in \Theta (I)} w_{k}^{\left( I, \theta \right)} p_{k}^{\left( \theta \right) }\!\left( x,\ell; I \right)  \label{eq:MGLMB_p} \\
	w_{k}^{\left( I, \theta \right)} & \propto &w_{k|k-1}\!\left( I\right)\left[ \overline{\psi}_{Z_{k}}^{(I, \theta)}(\cdot )\right] ^{I}  \label{eq:updateweight} \\
	\overline{\psi}_{Z_{k}}^{(I, \theta )}(\ell ) & = &\left\langle p_{k|k-1}\!\left(\cdot, \ell; I \right),\psi _{Z_{k}}(\cdot ,\ell ;\theta)\right\rangle \\
	p_{k}^{\left( \theta \right) }\left( x, \ell; I \right) & = &\frac{p_{k|k-1}(x,\ell; I )\psi _{Z_{k}}(x,\ell ;\theta )}{\overline{\psi}_{Z_{k}}^{( I, \theta )}(\ell )} \, .
	\label{eq58}
\end{IEEEeqnarray}

Note that the exact multi-object posterior density is not a M$\delta $-GLMB,
but a $\delta $-GLMB. The M$\delta $-GLMB update approximates the posterior
density by a M$\delta $-GLMB that preserves the posterior PHD and
cardinality distribution.

A tractable suboptimal multi-object estimate can be obtained from the
posterior M$\delta $-GLMB $\boldsymbol{\pi }_{k}=\{(w_{k}\!\left( I\right)
,p_{k}\!\left( \cdot ;I\right) )\}_{I\in \mathcal{F}\!\left( \mathbb{L}%
\right) }$ as follows: first determine the maximum a-posteriori cardinality
estimate $N^{\ast }$ from
\begin{equation}
\Pr (\left\vert X\right\vert =n)=\sum_{I\in \mathcal{F}\!\left( \mathbb{L}%
\right) }\delta _{n}(\left\vert I\right\vert ) \, w_{k}\!\left( I\right) ;
\end{equation}%
then determine the label set $I^{\ast }$ with highest weight $w_{k}\!\left(
I^{\ast }\right) $ among those with cardinality $N^{\ast }$; and finally
determine the expected values of the kinematic states from $p_{k}\!\left(
\cdot ,\ell ;I^{\ast }\right) $, $\ell \in I^{\ast }$. Alternatively, one can
determine the set $I^{\ast }$ of labels with the $N^{\ast }$ highest
existence probabilities $\sum_{I\in \mathcal{F}(\mathbb{L})}1_{I}(\ell
)w_{k}\!\left( I\right) $; and then the expected values of the
kinematic states from $\sum_{I\in \mathcal{F}(\mathbb{L})}1_{I}(\ell
)p_{k}\!\left( \cdot ,\ell ;I\right) $, $\ell \in I^{\ast }$.

In addition to the generality of the tracking solution, the consideration of
label-dependent $P_{S}$ and $P_{D}$ are useful in some applications. For
instance, in live cell microscopy the survival probability of a cell is also a function of its age. 
This can be accommodated by a label-dependent $P_{S}$ because the label contains the time of birth
and the age of a labeled state can be determined by subtracting the time of
birth from the current time. In some trackers, a track is considered to begin
when it is first detected, by convention. In this case, label-dependent $P_{D}$ would be able to capture this assumption, since the label provides
the time of birth.

\subsection{Consensus M$\protect\delta $-GLMB Filter}

\label{ssec:cmdglmbf} This subsection details the\textit{\ Consensus }M$%
\delta $-GLMB filter using a Gaussian mixture\ implementation. Each node $%
i\in \mathcal{N}$ of the network operates autonomously at each sampling
interval $k$, starting from its own previous estimates of the multi-object
distribution $\boldsymbol{\pi }^{(i)}$, with PDFs
$p\!\left( \cdot, \ell; I \right)$, $\forall \,\ell \in I$, $I\in \mathcal{F}%
\!\left( \mathbb{L}\right) $, represented by Gaussian mixtures, and
producing, at the end of $N$ consensus iterations, its new consensus
multi-object distribution $\boldsymbol{\pi }^{(i)}= \boldsymbol{\pi }_{N}^{(i)}$.

The steps of the Consensus M$\delta $-GLMB filter over the network $\mathcal{%
N}$ are given as follows.

\begin{enumerate}
\item Each node $i\in \mathcal{N}$ locally performs an M$\delta $-GLMB
prediction and update. The details of these two procedures are described in
the previous subsections.

\item At each consensus step, node $i$ transmits its data to neighbouring
nodes $j\in \mathcal{N}^{(i)}\backslash \{i\}$. Upon receiving data from its
neighbours, node $i$ carries out the fusion rule of Proposition \ref%
{pro:mdglmb:fusion} over its in-neighbours $\mathcal{N}^{(i)}$, i.e.
performs (\ref{eq:consenus:alg}) using information from $\mathcal{N}^{(i)}$.
A merging operation for each of the PDFs is applied to reduce the joint
communication-computation burden for the next consensus step. This procedure
is repeatedly applied for a chosen number $N\geq 1$ of consensus steps.

\item After consensus, an estimate of the multi-object state is obtained via
the procedure described in Table \ref{alg:mdglmb:estextr}.
\end{enumerate}

The operations executed locally by each node $i\in \mathcal{N}$ of the
network are summarized in Table \ref{alg:cmdglmb}.

\begin{table}[h!]
\caption{Consensus Marginalized $\protect\delta$-GLMB Filter}
\label{alg:cmdglmb}\renewcommand{\arraystretch}{1.3} \hrule\vspace{1mm}
\begin{algorithmic}[0]
	\Procedure{Consensus M$\delta$-GLMB}{\textsc{Node} $i$, \textsc{Time} $k$}
		\State \textsc{Local Prediction} \Comment{See subsection \ref{ssec:mdglmb:pred}}\vspace{0.5em}
		\State \textsc{Local Update} \Comment{See subsection \ref{ssec:mdglmb:corr}}\vspace{0.5em}
		\State \textsc{Marginalization} \Comment{See eqs. (\ref{eq:MGLMB_w})-(\ref{eq58})}\vspace{0.5em}
		\For{$n = 1, \dots, N$}\vspace{0.5em}
			\State \textsc{Information Exchange}\vspace{0.5em}
			\State \textsc{Fusion over $\mathcal{N}^{(i)}$} \Comment{See eqs. (\ref{eq:mdglmbfusion:w}) and  (\ref{eq:mdglmbfusion:pdf})}\vspace{0.5em}
			\State \textsc{Merging} \Comment{See \cite[Table II, Section III.C]{phd}}\vspace{0.5em}
		\EndFor\vspace{0.5em}
		\State \textsc{Estimate Extraction} \Comment{See algorithm in Table \ref{alg:mdglmb:estextr}}
	\EndProcedure\vspace{1mm}
\end{algorithmic}
\hrule
\end{table}
\begin{table}[h]
\caption{M$\protect\delta$-GLMB Estimate Extraction}
\label{alg:mdglmb:estextr}\renewcommand{\arraystretch}{1.3} \hrule\vspace{1mm}
\begin{algorithmic}[0]
		\State \textbf{\textsc{Input:}} $\boldsymbol{\pi }_{k}=\{(w_{k}\!\left( I \right), p_{k}\!\left(\cdot; I \right))\}_{I\in \mathcal{F}\!\left( \mathbb{L}\right) }$
		\State \textbf{\textsc{Output:}} $\mathbf{X}^{\ast}$\vspace{1mm}
	\end{algorithmic}
\hrule\vspace{1mm}
\begin{algorithmic}[0]
		\For{$c = 1, \dots, $}\vspace{0.5em}
			\State $\displaystyle{\rho(c) = \sum_{I \in \mathcal{F}(\mathbb{L})} \delta _{c}(\left\vert I\right\vert ) \, w_{k}\!\left( I \right)}$
		\EndFor\vspace{0.5em}
		\State $\displaystyle{N^{\ast} = \arg \max_{c} \rho(c)}$
		\State $\displaystyle I^{\ast} = \arg \max_{I \in \mathcal{F}_{N^{\ast}}\!\left( \mathbb{L} \right)} w_{k}^{(I)}$\vspace{0.5em}
		\State $\displaystyle{\mathbf{X}^{\ast} = \left\{ \left( x^{\ast}, \ell^{\ast} \right)\!: \ell^{\ast} \in I^{\ast},\, x^{\ast} = \arg \max_{x}p_{k}\!\left( x, \ell^{\ast}; I^{\ast} \right) \right\}}$
\end{algorithmic}
\hrulefill
\end{table}

We point out that in the algorithm of Table \ref{alg:cmdglmb}, each single object state lives in the space $\mathbb{X}\times \mathbb{L}_{0:k}$ and the multi-object state space at time $k$, $\mathcal{F}(\mathbb{X}\times \mathbb{L}_{0:k})$, is the same for all nodes. This is fully consistent with the fusion rule of Theorem 1 and the fusion rule for M$\protect\delta$-GLMBs.

In implementation,
each component of the M$\protect\delta$-GLMB, also known as a hypothesis, is indexed by an element of
$\mathcal{F}( \mathbb{L}_{0:k})$, i.e., a set of labels.
Since the cardinality of the label space $\mathbb{L}_{0:k}$ increases with time, each node performs
a pruning of the hypotheses (for instance by removing those with low weights so that the total number
of hypotheses never exceed a fixed number $I_{\rm max}$).
Hence, at time $k$ each node $i$ has a density containing at most $I_{\rm max}$ components with indices in $\mathbb I^{(i)}_k \subseteq \mathcal{F}( \mathbb{L}_{0:k})$,
and  the weights of the remaining components, i.e. those in $\mathcal{F}( \mathbb{L}_{0:k}) \setminus \mathbb I^{(i)}_k$, are set to zero.
As a result, when the densities of nodes $i$ and $j$ are fused, only the common components, i.e those belonging to
$\mathbb I^{(i)}_k \cap \mathbb I^{(j)}_k$ have non-zero weights.

\begin{rem}
\label{rem5} In \cite{FantacciPapi16} it has been shown that the centralized M$\delta$-GLMB filter features linear complexity in the number of sensors. As far as consensus M$\delta$-GLMB is concerned, each node has to carry out local prediction and local update, whose computational complexity is clearly independent of the number of nodes (sensors), and the consensus task (i.e. repeated KLA fusion over the subset of in-neighbors) which requires in the order of $N \, I_{max} \, d$ computations, $d$ being the node in-degree. 
\end{rem}

\subsection{The LMB Filter}

\label{ssec:lmbf} As suggested by its name, the LMB filter propagates an LMB
multi-object posterior density forward in time \cite{lmbf}. It is an
approximation of the $\delta $-GLMB filter \cite{vovo1,vovo2}.

\subsubsection{LMB Prediction}

\label{ssec:lmb:pred} Given the LMB multi-object posterior density $%
\boldsymbol{\pi }_{k-1}=\{(r_{k-1}\!\left( \ell \right), p_{k-1}\!\left(
\cdot, \ell \right))\}_{\ell \in \mathbb{L}_{-}}$, the multi-object
prediction density is the LMB \cite{lmbf}
\begin{equation}
\boldsymbol{\pi }_{k|k-1}=\left\{ \left( r_{S}\!\left( \ell \right),
p_{S}\!\left( \cdot, \ell \right)\right) \right\} _{\ell \in \mathbb{L}%
_{-}}\cup \left\{ \left(r_{B}\!\left( \ell \right), p_{B}\!\left( \cdot,
\ell \right)\right) \right\} _{\ell \in \mathbb{B}}  \label{eq:LMBPrediction}
\end{equation}
where
\begin{IEEEeqnarray}{rCl}
	r_{S}\!\left( \ell \right) & = & \left\langle P_{S}(\cdot ,\ell ),p_{k-1}(\cdot ,\ell)\right\rangle r_{k-1}\!\left( \ell \right)  \label{eq:LMBupdate1} \\
	p_{S}\!\left( \cdot, \ell \right) & = & \dfrac{\left\langle P_{S}(\cdot ,\ell )f_{k|k-1}(x|\cdot,\ell ),p_{k-1}(\cdot ,\ell )\right\rangle }{\left\langle P_{S}(\cdot ,\ell),p_{k-1}(\cdot ,\ell )\right\rangle }   \label{eq:LMBupdate2}
\end{IEEEeqnarray}

\subsubsection{LMB Update}

\label{ssec:lmb:corr} Given the LMB multi-object prediction density $%
\boldsymbol{\pi }_{k|k-1}=\{(r_{k|k-1}\!\left( \ell \right),
p_{k|k-1}\!\left( \cdot, \ell \right))\}_{\ell \in \mathbb{L}}$, the LMB
updated density is given by $\boldsymbol{\pi}_{k}=\{(r_{k}\!\left( \ell
\right), p_{k}\!\left( \cdot, \ell \right))\}_{\ell \in \mathbb{L}}$, where
\begin{IEEEeqnarray}{rCl}
	r_{k}\!\left( \ell \right) & = & \sum_{(I,\theta )\in \mathcal{F}(\mathbb{L})\times \Theta (I)}\!\!\!\!1_{I}(\ell )w_{k}^{(\theta )}\!\left( I \right)  \label{eq:existenceBernoulli} \\
	p_{k}\!\left( x, \ell \right) & = & \frac{1}{r_{k}\!\left( \ell \right)}\!\sum_{(I,\theta)\in \mathcal{F}(\mathbb{L})\times \Theta (I)}\!\!\!\!1_{I}(\ell)w_{k}^{(\theta )}\!\left( I \right) p_{k}^{(\theta )\!}(x,\ell ) \label{eq:spatialBernoulli} \IEEEeqnarraynumspace\\
	w_{k}^{(\theta )\!}\!\left( I \right) & \propto & \left[ \overline{\psi}_{Z_{k}}^{(\theta )}\right] ^{I}\!\left[ 1-r_{k|k-1}\!\left( \cdot \right)\right]^{\mathbb{L}\backslash I}\!\left[ 1_{\mathbb{L}} \, r_{k|k-1} \right]^{I}  \\
	p_{k}^{(\theta )}(x,\ell ) & = & \frac{p_{k|k-1}(x,\ell ) \, \psi_{Z_{k}}(x,\ell ;\theta )}{\overline{\psi}_{Z_{k}}^{(\theta )}(\ell )} \\
	\overline{\psi}_{Z_{k}}^{(\theta )}(\ell ) & = & \left\langle p_{k|k-1}(\cdot ,\ell ),\psi _{Z_{k}}(\cdot ,\ell ;\theta )\right\rangle \,.
\end{IEEEeqnarray}

Note that the exact multi-object posterior density is not an LMB, but a
GLMB. The LMB update approximates the GLMB posterior by an LMB that matches
the unlabeled PHD. The reader is referred to \cite{lmbf} for an efficient
implementation of the LMB filter.

\subsection{Consensus LMB Filter}

\label{ssec:clmbf} This subsection describes the\textit{\ Consensus }LMB
filter using a Gaussian mixture\ implementation. The steps of the Consensus%
\textit{\ }LMB filter are the same as the Consensus M$\delta $-GLMB tracking
filter described in section \ref{ssec:cmdglmbf}, with the LMB prediction and
update in place of those of the M$\delta $-GLMB. Each node $i\in \mathcal{N}$
of the network operates autonomously at each sampling interval $k$, starting
from its own previous estimates of the multi-object distribution $%
\boldsymbol{\pi }^{(i)}$, with PDFs 
$p\!\left( \cdot, \ell \right)$, $\forall \,\ell \in \mathbb{L}$,
represented by Gaussian mixtures, and producing, at the end of $N$ consensus
iterations, its new consensus multi-object distribution $\boldsymbol{\pi }%
^{(i)}= \boldsymbol{\pi }_{N}^{(i)}$. The operations executed locally
by each node $i\in \mathcal{N}$ of the network are summarized in Table \ref%
{alg:clmb}.

\begin{table}[h]
\caption{Consensus LMB Filter}
\label{alg:clmb}\renewcommand{\arraystretch}{1.3} \hrule\vspace{1mm}
\begin{algorithmic}[0]
\Procedure{Consensus LMB}{\textsc{Node} $i$, \textsc{Time} $k$}
		\State \textsc{Local Prediction} \Comment{See subsection \ref{ssec:lmb:pred}}\vspace{0.5em}
		\State \textsc{Local Update} \Comment{See subsection \ref{ssec:lmb:corr}}\vspace{0.5em}
		\For{$n = 1, \dots, N$}\vspace{0.5em}
			\State \textsc{Information Exchange}\vspace{0.5em}
			\State \textsc{Fusion over $\mathcal{N}^{(i)}$} \Comment{See eqs. (\ref{eq:lmbfusion:ex}) and (\ref{eq:lmbfusion:pdf})}\vspace{0.5em}
			\State \textsc{Merging} \Comment{See \cite[Table II, Section III.C]{phd}}\vspace{0.5em}
		\EndFor\vspace{0.5em}
		\State \textsc{Estimate Extraction} \Comment{See algorithm in Table \ref{alg:clmb:estextr}}
\EndProcedure\vspace{1mm}
\end{algorithmic}
\hrule
\end{table}
\begin{table}[h]
\caption{LMB Estimate Extraction}
\label{alg:clmb:estextr}\renewcommand{\arraystretch}{1.3} \hrule\vspace{1mm}
\begin{algorithmic}[0]
	\State \textbf{\textsc{Input:}} $\boldsymbol{\pi}_{k} = \left\{ r_{k}\!\left( \ell \right), p_{k}\!\left( \cdot \right) \right\}_{\ell \in \mathbb{L}}$, $N^{\ast}$
	\State \textbf{\textsc{Output:}} $\mathbf{X}^{\ast}$\vspace{1mm}
\end{algorithmic}
\hrule\vspace{1mm}
\begin{algorithmic}[0]
\For{$c = 1, \dots, N^{\ast}$}\vspace{0.5em}
	\State $\displaystyle{\rho(c) = \sum_{I \in \mathcal{F}(\mathbb{L})} \delta _{c}(\left\vert I\right\vert ) \prod_{\ell \in \mathbb{L}\backslash I}\left( 1 - r_{k}\!\left( \ell \right) \right) \prod_{\ell \in I}r_{k}\!\left( \ell \right)}$
\EndFor\vspace{0.5em}
\State $\displaystyle{C^{\ast} = \arg \max_{c} \rho(c)}$
\State $\hat{\mathbb{L}} = \varnothing$\vspace{0.5em}
\For{$c^{\ast} = 1, \dots, C^{\ast}$}\vspace{0.5em}
	\State $\displaystyle{\mathbb{L}^{\ast} = \mathbb{L}^{\ast} \cup \arg \max_{\ell \in \mathbb{L}\backslash\mathbb{L}^{\ast}} r_{k}\!\left( \ell \right)}$
\EndFor\vspace{0.5em}
\State $\displaystyle{\mathbf{X}^{\ast} = \left\{ \left( x^{\ast}, \ell^{\ast} \right) \!: \ell^{\ast} \in \mathbb{L}^{\ast},\, x^{\ast} = \arg \max_{x} p_{k}\!\left( x, \ell^{\ast} \right) \right\}}$
\end{algorithmic}
\hrulefill
\end{table}

Similar to M$\delta$-GLMB Consensus, each node performs pruning of the components (or hypotheses) so as to cap their numbers.
For an LMB, each component is indexed by an element of $\mathbb L^{(i)}_k$, i.e. a label.
Hence, for LMB Consensus, at time $k$ each node $i$  has a density containing at most $I_{\rm max}$  components
indexed by $\mathbb L^{(i)}_k \subseteq \mathbb L_{0:k}$,
and the weights of those components indexed by $\mathbb L_{0:k} \setminus \mathbb L^{(i)}_k $
are set to zero.
Table \ref{tab:borderlineinit}  summarizes  the information exchanged among the nodes for both M$\protect\delta$-GLMB and LMB trackers.

 \begin{table}[h]
\caption{Information exchanged at time interval $k$ by agents $i\in \mathcal{%
N}$ for both M$\protect\delta $-GLMB and LMB trackers. It is assumed that
the exchanged information are represented with 4 bytes floating point
variables.}
\label{tab:borderlineinit}\centering
\renewcommand{\arraystretch}{1.3} \setlength\arrayrulewidth{0.5pt}  \arrayrulecolor{black}  \setlength\doublerulesep{0.5pt}  \doublerulesepcolor{black}
\scalebox{.95}{
		\begin{tabular}{|c|c|c|}
			\hline
			\textbf{Tracker} & \textbf{Information exchanged} & \textbf{Total bytes exchanged}\\
			\hline
			M$\delta$-GLMB & $\forall ~ I \in \mathbb I^{(i)}_k$: $\omega_{k}^{(i)}$, $p_{k}^{(i)}(x, \ell; I)$ & $ 4 \displaystyle{ \sum_{I \in \mathbb I^{(i)}_k}}(1 + (4 + 10)|I|)$\\
			\hline
			LMB & $\forall ~ \ell \in \mathbb L^{(i)}_k $: $r_{k}^{(i)}(\ell)$, $p_{k}^{(i)}(x, \ell)$ & $ 4 (1 + (4 + 10)|\mathbb L^{(i)}_k|)$\\
			\hline
		\end{tabular}
		}
\end{table}

\section{Performance evaluation}

\label{sec:performance} 
\begin{figure}[h]
	\centering
	\includegraphics[width=0.5\columnwidth]{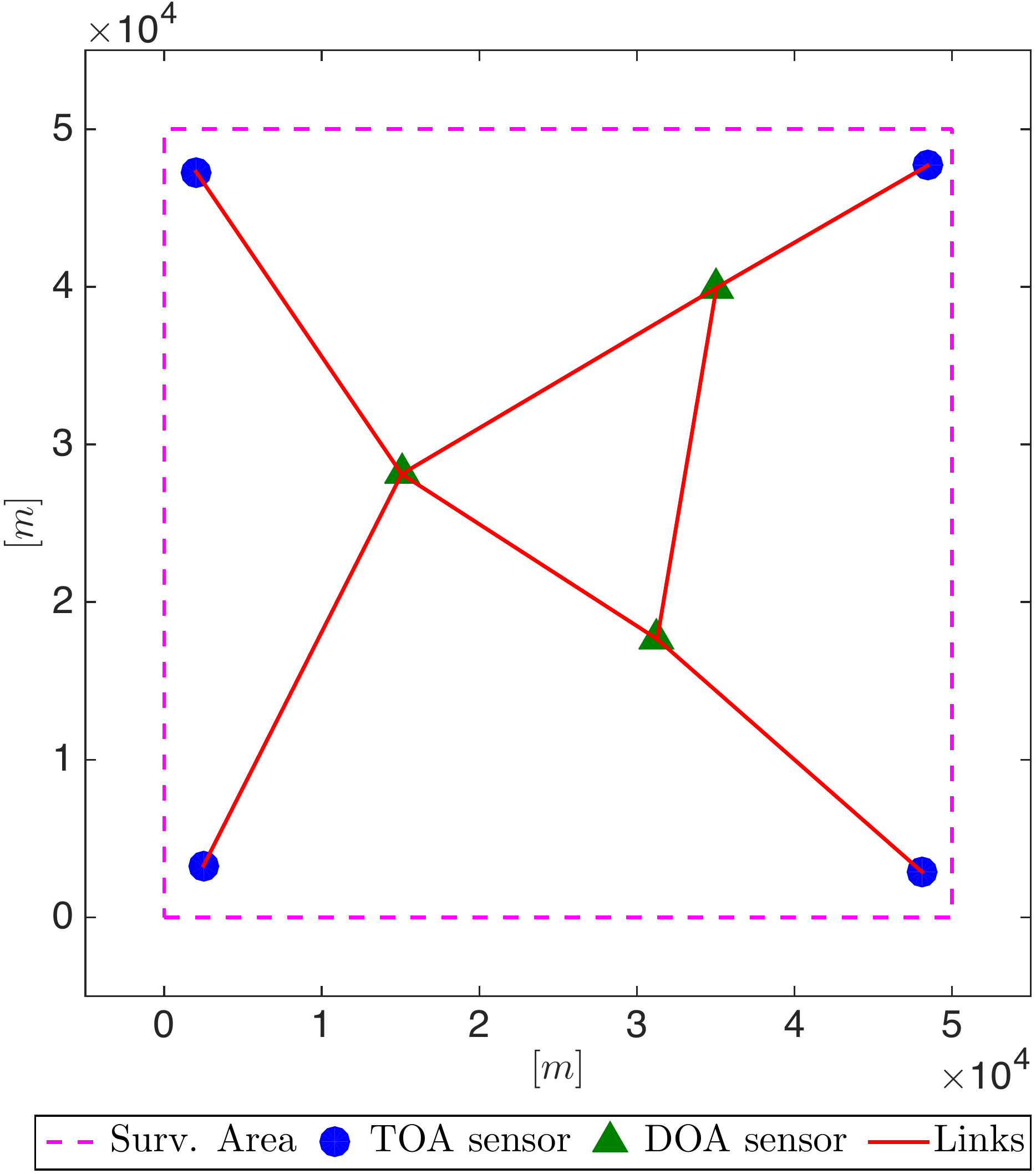}\vspace{-1em}
	\caption{Network with 7 sensors: 4 TOA and 3 DOA.}
	\label{fig:4toa3doa}
\end{figure}

\begin{figure}[h]
	\centering
	\includegraphics[width=0.5\columnwidth]{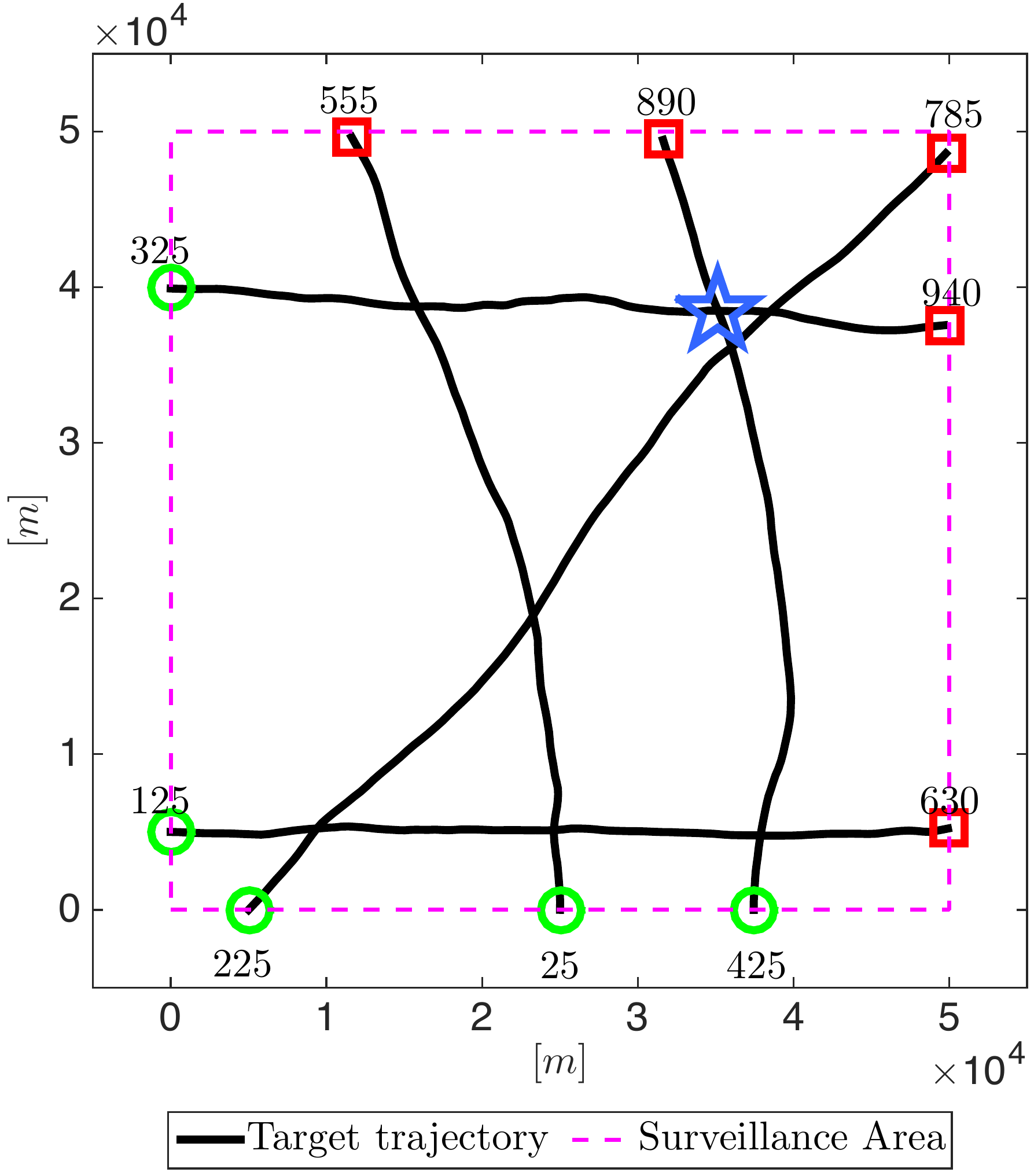}\vspace{-1em}
	\caption{Target trajectories considered in the simulation experiment. The start/end point for each trajectory is denoted, respectively, by $\bullet\backslash\blacksquare$. The {\Large$\star$} indicates a rendezvous point.}
	\label{fig:5trajectories}
\end{figure}

To assess the performance of the proposed consensus
multi-object tracking filters, we consider a $2$-D multi-object tracking
scenario over a surveillance area of $50\times 50\,km^{2}$, wherein the
sensor network depicted in Fig. \ref{fig:4toa3doa} is deployed. The scenario
consists of $5$ objects as shown in Fig. \ref{fig:5trajectories}.
The proposed trackers are also compared with the Consensus CPHD filter of \cite{ccphd} which, however, does not provide tracks, and with the \textit{centralized} M$\delta$-GLMB filter \cite{mdglmbf} which recursively processes all the measurements collected by the nodes, thus providing a performance reference for the distributed filters.

The kinematic object state is denoted by $x=\left[ p_{x},\,\dot{p}_{x}%
,\,p_{y},\,\dot{p}_{y}\right] ^{\top }$, i.e. the planar position and
velocity. The motion of objects is modeled by the filters according to the
Nearly-Constant Velocity (NCV) model \cite%
{farina1,farina2,barshalom1,barshalom2}: $\mathcal{N}\!\left(
x_{k};Fx_{k-1},Q\right) $, where
\begin{equation*}
F=\left[
\begin{array}{cccc}
1 & T_{s} & 0 & 0 \\
0 & 1 & 0 & 0 \\
0 & 0 & 1 & T_{s} \\
0 & 0 & 0 & 1%
\end{array}%
\right],  ~Q=\sigma _{w}^{2}\!\!\left[
\begin{array}{cccc}
\frac{T_{s}^{4}}{4} & \frac{T_{s}^{3}}{2} & 0 & 0 \\
\frac{T_{s}^{3}}{2} & T_{s}^{2} & 0 & 0 \\
0 & 0 & \frac{T_{s}^{4}}{4} & \frac{T_{s}^{3}}{2} \\
0 & 0 & \frac{T_{s}^{3}}{2} & T_{s}^{2}%
\end{array}%
\right] \!,
\end{equation*}%
$\sigma _{w}=5\,m/s^{2}$ and the sampling interval is $T_{s}=5\,s$.
Objects pass through the surveillance area and partial prior information of the object birth locations is assumed.
Accordingly, a $10$-component LMB RFS $\boldsymbol{\pi}_{B}=\{(r_{B}\!\left( \ell \right),p_{B}\!\left( \cdot, \ell \right))\}_{\ell \in \mathbb{B}}$ is used to model the birth process. Table \ref{tab:borderlineinit} gives a detailed summary of such components.
The aim of using such a birth process is to cover all possible locations where objects appear, but also locations where no objects are present or born. In this way, it is also tested the algorithm’s ability of ruling out possible false objects, arising in wrong birth locations, that are generated by clutter measurements.
\begin{table}[h]
\label{tab:borderlineinit}
\caption{Components of the LMB RFS birth process at a given time $k$}%
\renewcommand{\arraystretch}{1.3} \setlength\arrayrulewidth{0.5pt}%
\arrayrulecolor{black} \setlength\doublerulesep{0.5pt}%
\doublerulesepcolor{black}
\centering
\vspace{-2.85em}
\begin{equation*}
	r\!\left( \ell \right) = 0.09 \mbox{, } p_{B}\!\left( x, \ell \right) = \mathcal{N}\!\left( x;\, m_{B}\!\left( \ell \right), P_{B} \right)
\end{equation*}
\begin{equation*}
	P_{B} = \operatorname{diag}\!\left( 10^{6}, 10^{4}, 10^{6}, 10^{4} \right)
\end{equation*}
\scalebox{1}{
	\begin{tabular}{|c|c|}
		\hline
		$\ell$ & $m_{B}\!\left( \ell \right)$\\
		\hline
		$\left( k, \, 1 \right)$ & $\left[ 0, \, 0, \, 40000,\, 0 \right]^{\top}$ \\
		$\left( k, \, 2 \right)$ & $\left[ 0, \, 0, \, 25000,\, 0 \right]^{\top}$ \\
		$\left( k, \, 3 \right)$ & $\left[ 0, \, 0, \, 5000,\, 0 \right]^{\top}$ \\
		$\left( k, \, 4 \right)$ & $\left[ 5000, \, 0, \, 0,\, 0 \right]^{\top}$ \\
		$\left( k, \, 5 \right)$ & $\left[ 25000, \, 0, \, 0,\, 0 \right]^{\top}$ \\
		$\left( k, \, 6 \right)$ & $\left[ 36000, \, 0, \, 0,\, 0 \right]^{\top}$ \\
		$\left( k, \, 7 \right)$ & $\left[ 50000, \, 0, \, 15000,\, 0 \right]^{\top}$ \\
		$\left( k, \, 8 \right)$ & $\left[ 50000, \, 0, \, 40000,\, 0 \right]^{\top}$ \\
		$\left( k, \, 9 \right)$ & $\left[ 40000, \, 0, \, 50000,\, 0 \right]^{\top}$ \\
		$\left( k, \, 10 \right)$ & $\left[ 10000, \, 0, \, 50000,\, 0 \right]^{\top}$ \\
		\hline
	\end{tabular}
	}
\end{table}

The sensor network considered in this example (see Fig. \ref{fig:4toa3doa})
consists of $4$ \textit{range-only} (Time Of Arrival, TOA) and $3$ \textit{%
bearing-only} (Direction Of Arrival, DOA) sensors characterized by the
following measurement functions:
\begin{equation*}
\begin{array}{c}
h^{(i)}(x)=\left\{
\begin{array}{ll}
\angle \lbrack \left( p_{x}-x^{(i)}\right) +j\left( p_{y}-y^{(i)}\right) ],
& \mbox{DOA} \\[0.5em]
\sqrt{\left( p_{x}-x^{(i)}\right) ^{2}+\left( p_{y}-y^{(i)}\right) ^{2}}, & %
\mbox{TOA}%
\end{array}%
\right.%
\end{array}%
\end{equation*}%
where $(x^{(i)},y^{(i)})$ represents the known position of sensor $i$. The
standard deviation of DOA and TOA measurement noises are taken respectively
as $\sigma _{DOA}=1^{\circ}$ and $\sigma _{TOA}=100\,m$. Each sensor has a
uniform clutter spatial distribution over the surveillance area. Due to the
non linearity of the sensor models, the \textit{Unscented Kalman Filter}
(UKF) \cite{juluhl2004} is used to update means and covariances of the
Gaussian mixture components.

Three different scenarios with various Poisson clutter rates $\lambda _{c}$
and constant detection probabilities $P_{D}$ are considered:

\begin{itemize}
\item \textbf{High SNR}: $\lambda_{c} = 5$, $P_{D} = 0.99$. These parameters
were used in \cite{ccphd} and, therefore, will be used as a first
comparison test.

\item \textbf{Low SNR}: $\lambda _{c}=15$, $P_{D}=0.99$. These parameters
emulate a realistic a scenario characterized by high clutter rate.

\item \textbf{Low $\mathbf{P_{D}}$}: $\lambda _{c}=5$, $P_{D}=0.7$. These
parameters test the distributed algorithms in the presence of severe
misdetection.
\end{itemize}

Multi-object tracking performance is evaluated in terms of the \textit{%
Optimal SubPattern Assignment} (OSPA) metric \cite{schvovo2008} with
Euclidean distance, i.e. $p=2$, and cutoff $c=600\,m$. The reported metric is
averaged over $100$ Monte Carlo trials for the same object trajectories but
different, independently generated, clutter and measurement noise
realizations. The duration of each simulation trial is fixed to $1000\,s$ ($%
200 $ samples).

The Consensus M$\delta $-GLMB and the Consensus LMB filters are limited to $3000$ hypotheses and are coupled with the \textit{parallel CPHD look ahead strategy} described in \cite{vovo1,vovo2}. The CPHD filter is similarly limited to the same number of components through pruning and merging of mixture components \cite{cphd}.

The parameter setting used in \cite{ccphd} for the Consensus CPHD filter has been adopted for the present simulation campaigns. In particular, the survival probability is $P_{S}=0.99$; the maximum number of Gaussian components is $N_{max}=25$; the merging threshold is $\gamma _{m}=4$; the truncation threshold is $\gamma _{t}=10^{-4}$; the extraction threshold is $\gamma _{e}=0.5$; the birth intensity function is the PHD of the LMB RFS of Table \ref{tab:borderlineinit}.

$N=1$ and $N=3$ consensus steps have been considered for the simulations. The choice $N=1$ is clearly the most critical one for tracking performance due to the minimal amount of information exchanged during consensus, but at the same time the most parsimonious in terms of data communication load. On the other hand, the choice $N=3$ (the diameter of the network, i.e. the maximum distance between any two nodes in the network) allows to show the benefits of performing multiple consensus steps in terms of performance gain.
The case $N=1$ is certainly the most interesting one for comparing the various multi-object consensus filters as it highlights the capability of the proposed fusion technique to provide satisfactory results with little information exchanged and fused; the case $N=3$ is interesting to understand how many consensus steps are needed to achieve comparable performance to the centralized setting (i.e. the M$\delta$-GLMB filter) where measurements from all the sensors are recursively processed by a single tracker.

\subsection{High SNR}

\begin{figure}[h]
	\begin{minipage}[t][][t]{\columnwidth}
		\centering
		\includegraphics[width=0.8\columnwidth]{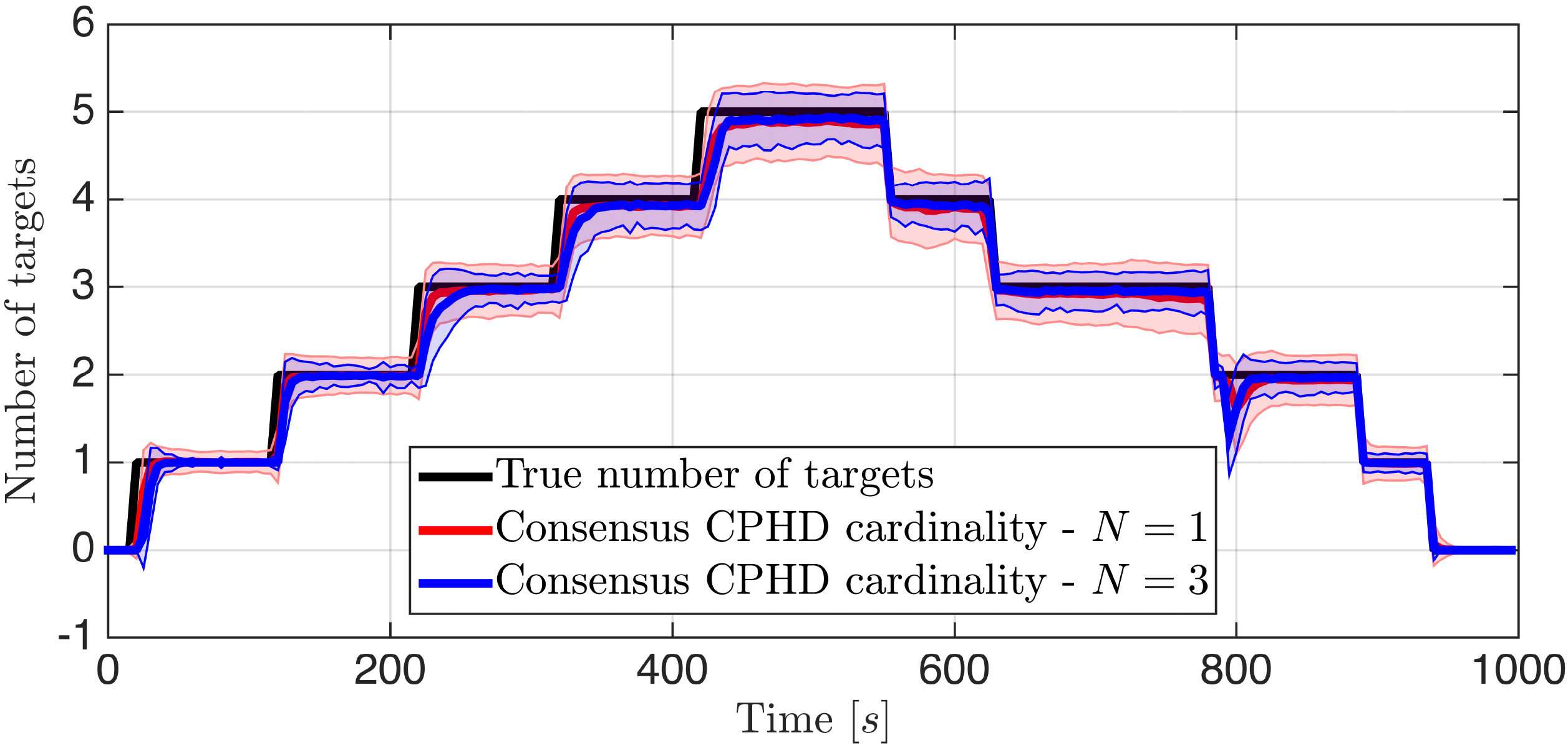}\vspace{-1em}
		\caption{Cardinality statistics for Consensus CPHD filter under high SNR.}
		\label{fig:1:cardCCPHD}
        \end{minipage}\vspace{1em}
	\begin{minipage}[t][][t]{\columnwidth}
		\centering
		\includegraphics[width=0.8\columnwidth]{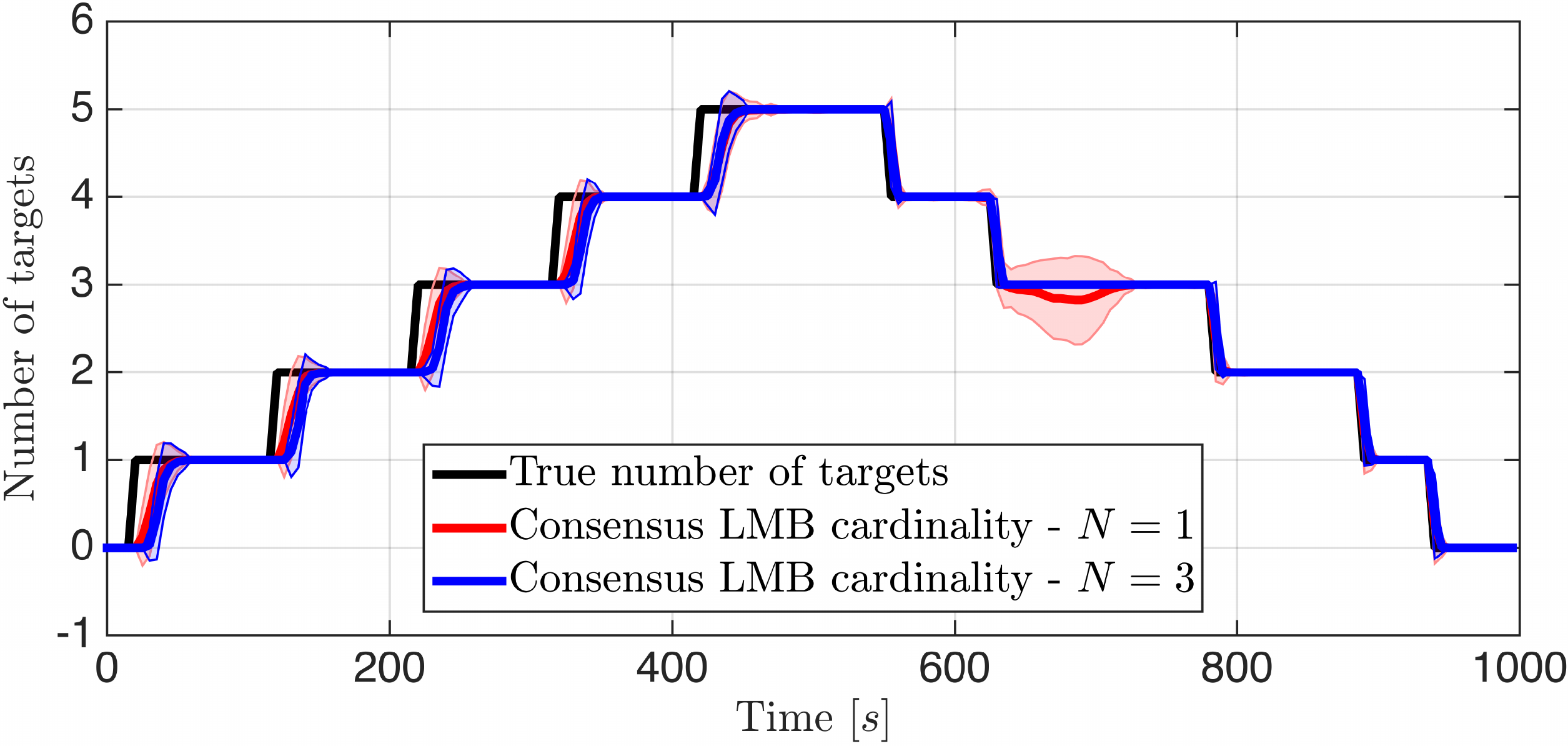}\vspace{-1em}
		\caption{Cardinality statistics for Consensus LMB tracker under high SNR.}
		\label{fig:1:cardCLMB}
        \end{minipage}\vspace{1em}
	\begin{minipage}[t][][t]{\columnwidth}
		\centering
		\includegraphics[width=0.8\columnwidth]{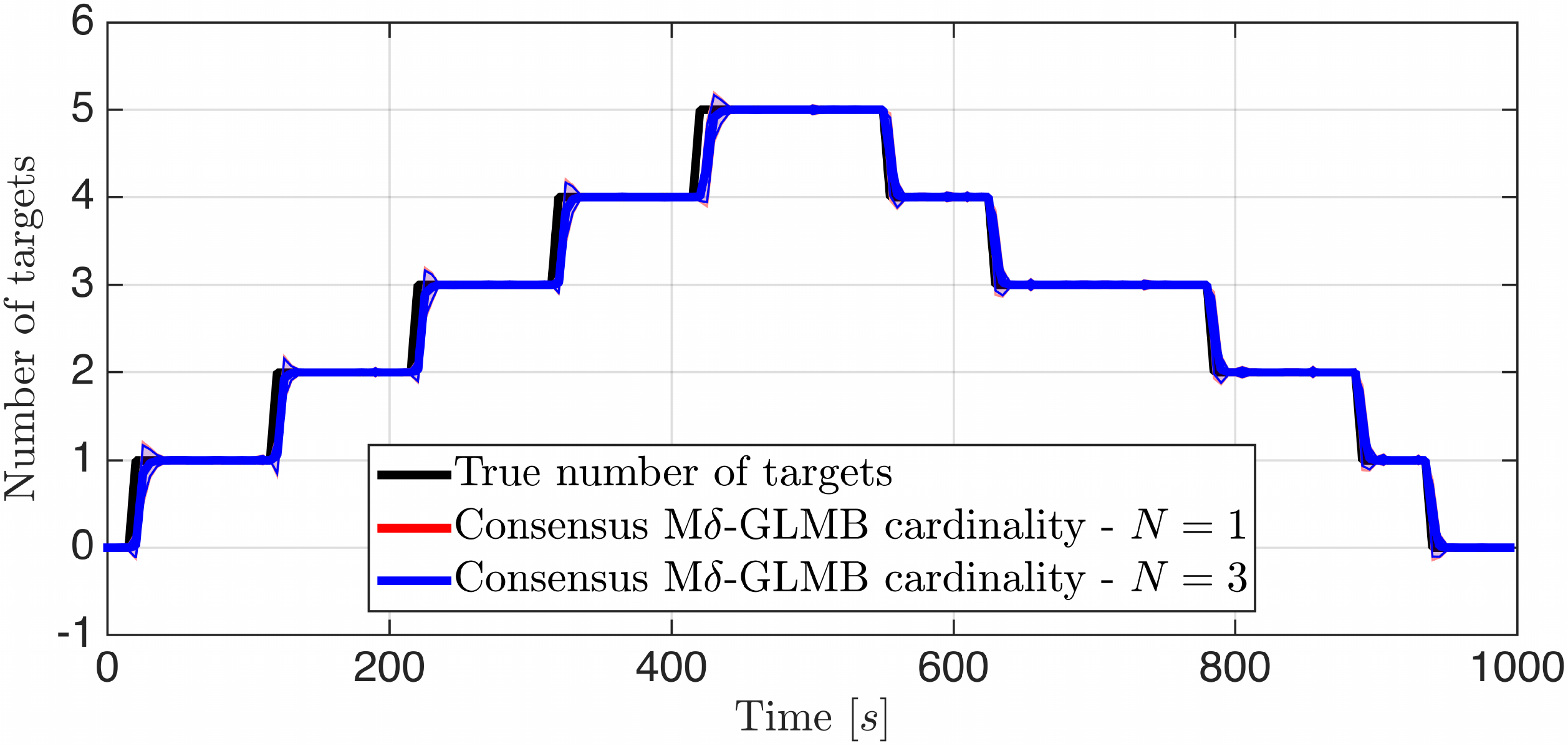}\vspace{-1em}
		\caption{Cardinality statistics for Consensus M$\delta$-GLMB tracker under high SNR.}
		\label{fig:1:cardCMDGLMB}
        \end{minipage}
\end{figure}

Figs. \ref{fig:1:cardCCPHD}, \ref{fig:1:cardCLMB} and \ref{fig:1:cardCMDGLMB}
display the statistics (mean and standard deviation) of the estimated number
of objects obtained, respectively, with the Consensus CPHD, the Consensus LMB and
the Consensus M$\delta $-GLMB filters. Observe that all three distributed
algorithms estimate the object cardinality accurately, with the Consensus M$%
\delta $-GLMB exhibiting the best cardinality estimate (with least
variance). Note that the difficulties introduced by the rendezvous point
(e.g. merged or lost tracks) are correctly addressed by all three
distributed algorithms.
Performing $N=3$ consensus steps provides a significant improvement only to the cardinality estimation of the Consensus CPHD filter.

\begin{figure}[h]
	\includegraphics[width=\columnwidth]{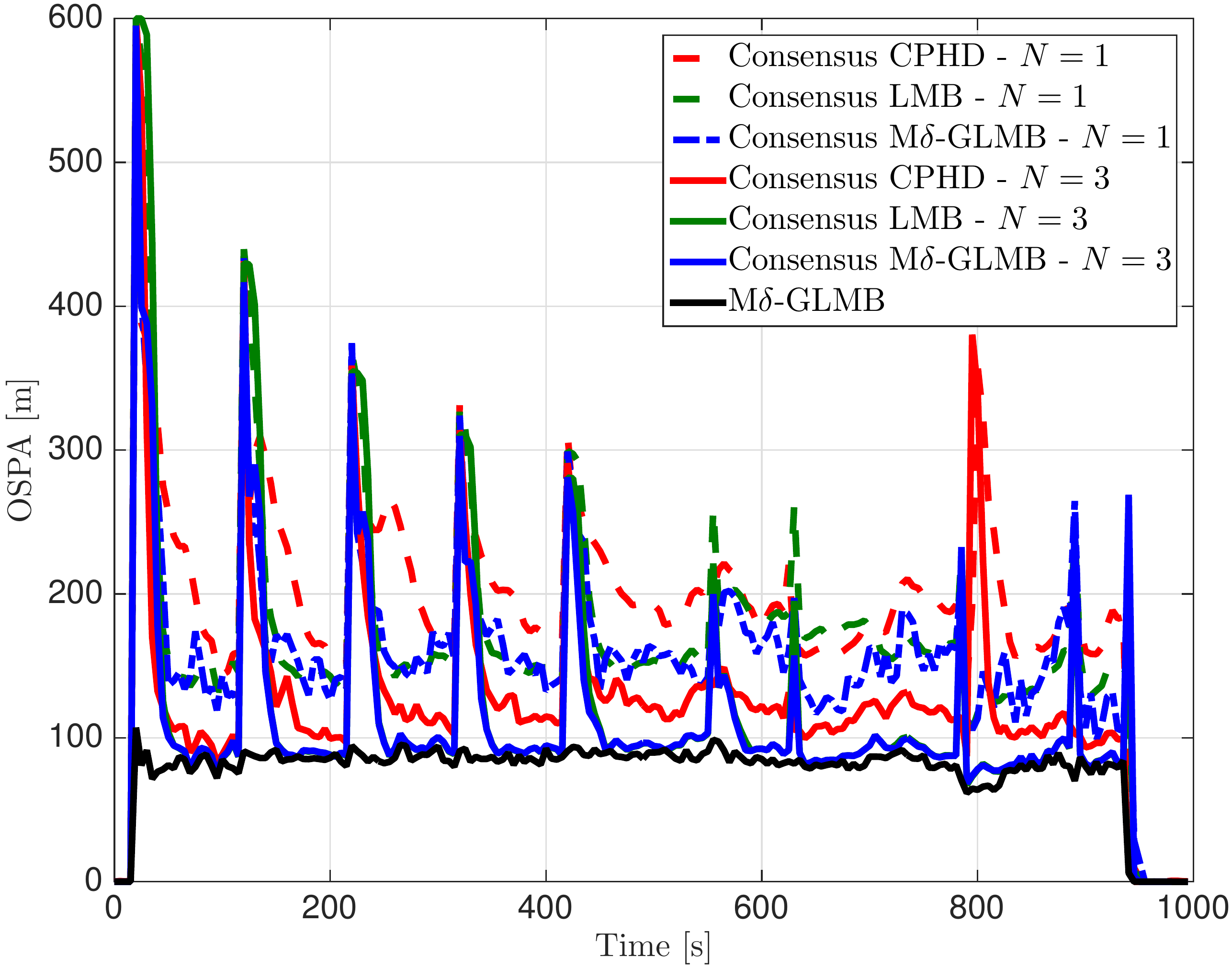}\vspace{-1em}
	\caption{OSPA distance ($c = 600 \, [m]$, $p = 2$) under high SNR.}
	\label{fig:1:ospa}
\end{figure}

Fig. \ref{fig:1:ospa} shows the OSPA distance for the three algorithms.
Compared to Consensus CPHD, the improved localization performance of the Consensus
LMB and the Consensus M$\delta $-GLMB is attributed to two factors: (a) the
\textquotedblleft spooky effect\textquotedblright\ \cite{spooky} causes the
Consensus CPHD filter to temporarily drop tracks which are subjected to missed
detections and to declare multiple estimates for existing tracks in place of
the dropped tracks, and (b) the two tracking filters are generally able to
better localize objects due to a more accurate propagation of the posterior
density. Note that Consensus LMB and Consensus M$\delta $-GLMB filters
exhibit similar performance since the additional approximation in the LMB
filter (see (\ref{eq:existenceBernoulli})-(\ref{eq:spatialBernoulli})) is
not significant in high SNR.
Multiple consensus steps provide a remarkable performance gain in terms of state estimation error. As it can be seen from Figs. \ref{fig:1:cardCCPHD}, \ref{fig:1:cardCLMB} and \ref{fig:1:cardCMDGLMB}, the cardinality is, on average, correctly estimated by all sensors in all Monte Carlo trials. Thus, the main contributor to the OSPA error reduction is the state estimation error. The object births and deaths are responsible for the peaks of the OSPA error in the distributed algorithms. The peaks are not present in the M$\delta$-GLMB because it makes use of all measurements provided by the sensors at each time interval $k$, while the distributed trackers only use the local measurements and require a few fusion steps in order to properly estimate the states of the objects. It is worth noticing that with $N=3$ the OSPA error of the Consensus LMB and Consensus M$\delta$-GLMB after each peak is very close to the one of the M$\delta$-GLMB.

\subsection{Low SNR}

\begin{figure}[h]
	\begin{minipage}[t][][t]{\columnwidth}
		\centering
		\includegraphics[width=0.8\columnwidth]{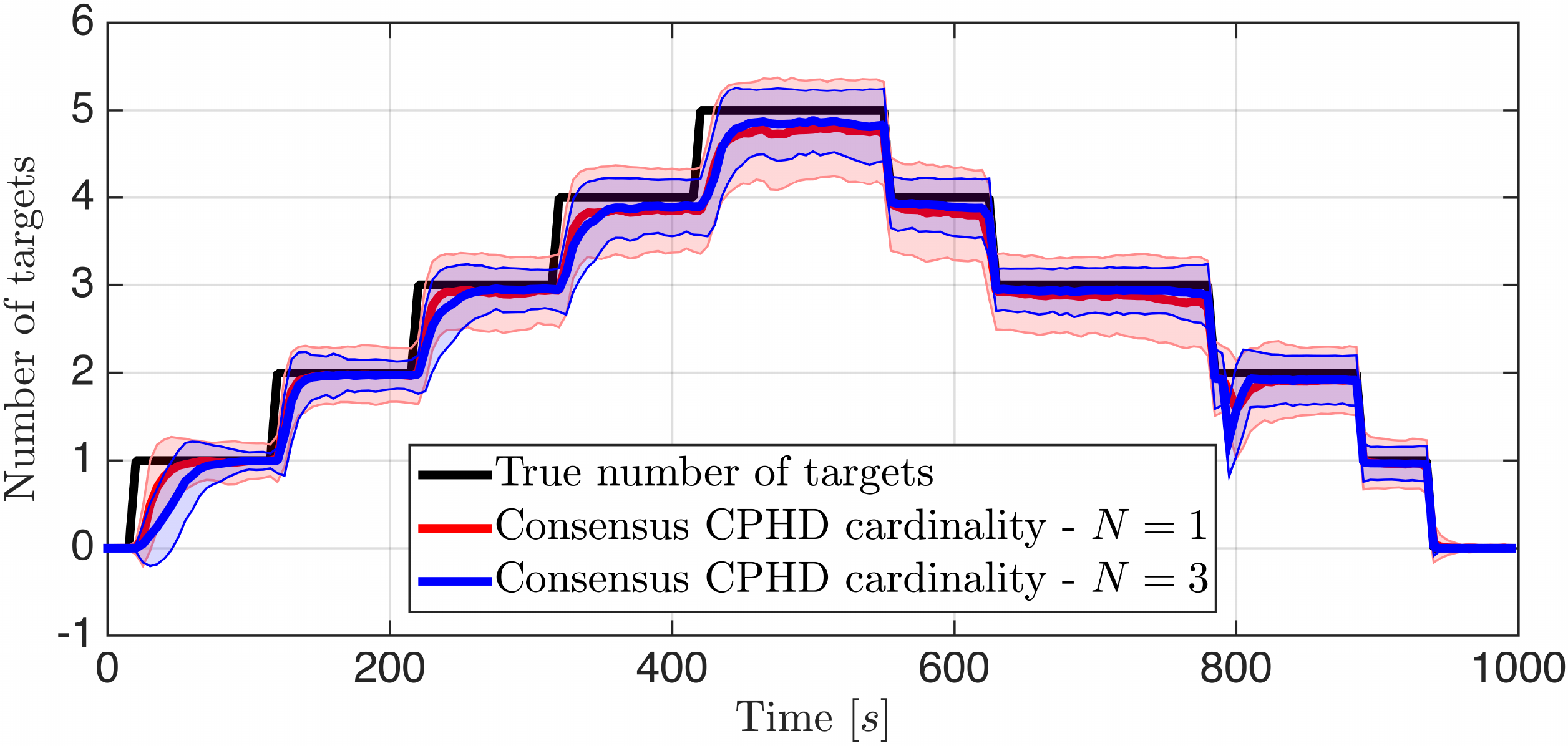}\vspace{-1em}
		\caption{Cardinality statistics for Consensus CPHD filter under low SNR.}
		\label{fig:2:cardCCPHD}
        \end{minipage}\vspace{1em}
	\begin{minipage}[t][][t]{\columnwidth}
		\centering
		\includegraphics[width=0.8\columnwidth]{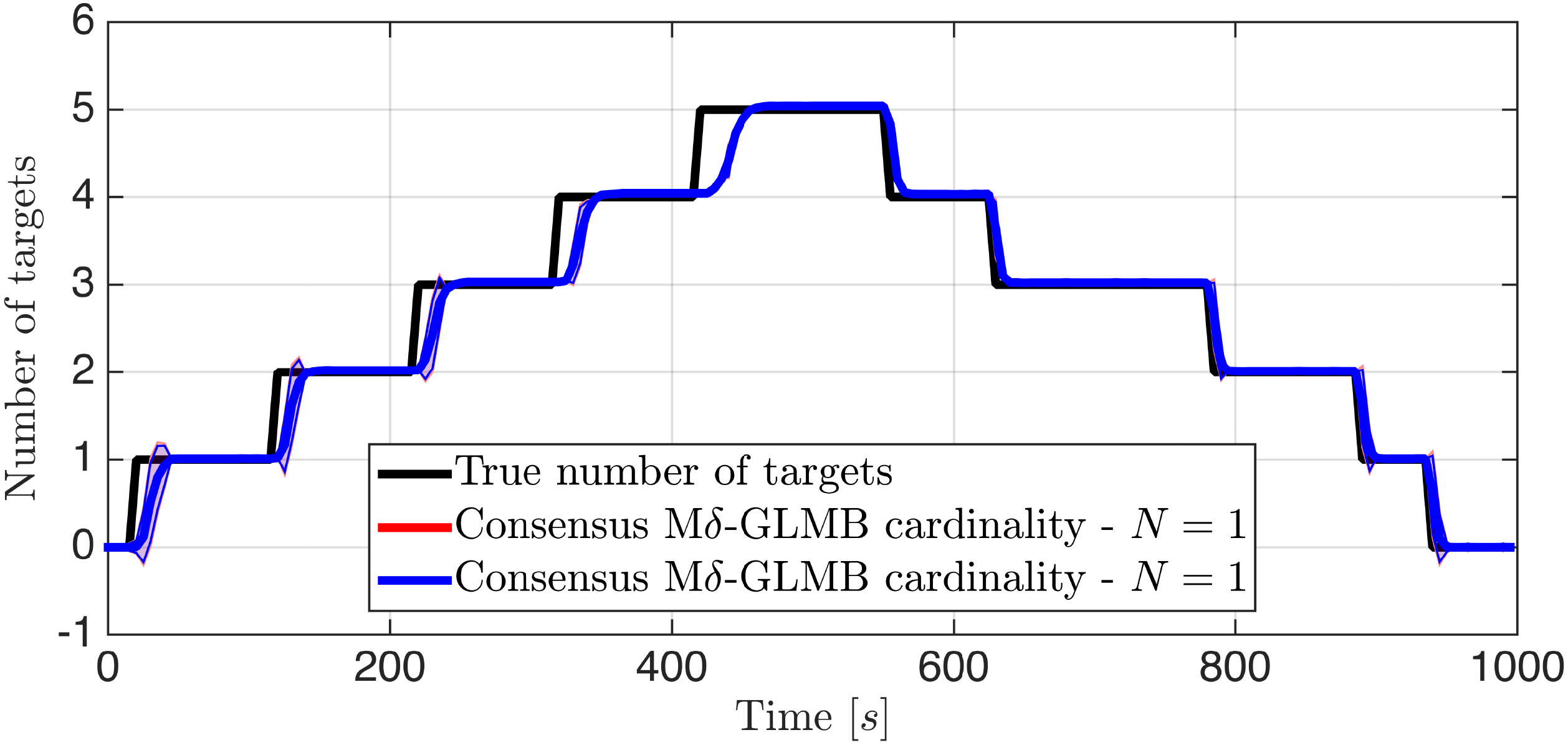}\vspace{-1em}
		\caption{Cardinality statistics for Consensus M$\delta$-GLMB tracker under low SNR.}
		\label{fig:2:cardCMDGLMB}
        \end{minipage}\vspace{1em}
\end{figure}

Figs. \ref{fig:2:cardCCPHD} and \ref{fig:2:cardCMDGLMB} display the
statistics (mean and standard deviation) of the estimated number of objects
obtained, respectively, with the Consensus CPHD and the Consensus M$\delta $-GLMB.
Observe that these two distributed filters estimate the object cardinality
accurately, with the Consensus M$\delta $-GLMB exhibiting again better
cardinality estimate (with lower variance).

Note that the Consensus LMB filter fails to track the objects. The problem
is due to the approximation of the GLMB posteriors by LMBs, which becomes
more severe with low SNR. In particular, each local tracker fails to
properly capture the existence probability of the tracks due to three main
factors: (a) no local observability, (b) high clutter rate and (c) loss of
the full posterior cardinality distribution after the LMB approximation.
Having low existence probabilities, the extraction of the tracks fails even if the single object densities are correctly propagated in time.

\begin{figure}[h]
	\begin{minipage}[t][][t]{\columnwidth}
		\centering
		\includegraphics[width=\columnwidth]{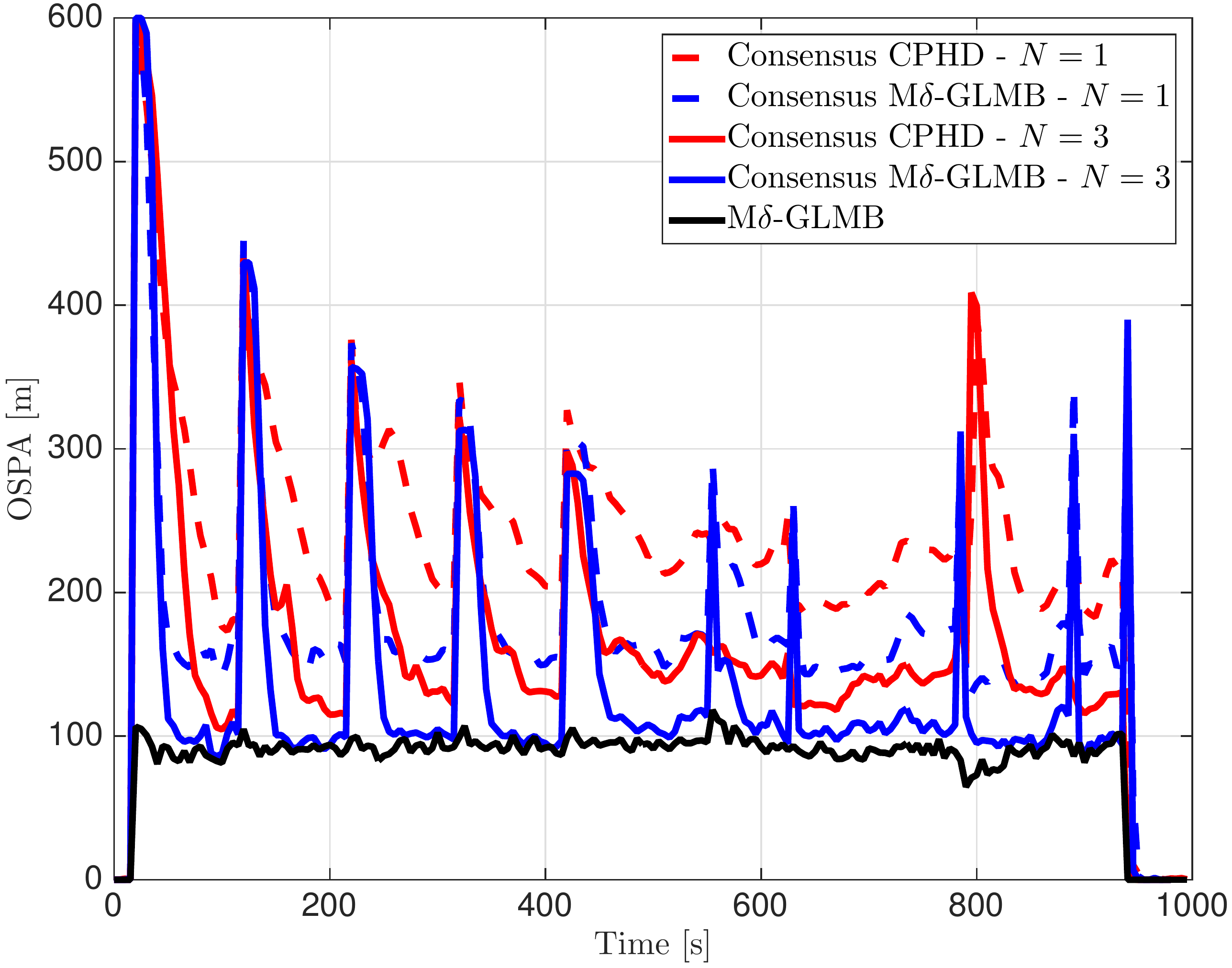}\vspace{-1em}
		\caption{OSPA distance ($c = 600 \, [m]$, $p = 2$) under low SNR.}
		\label{fig:2:ospa}
        \end{minipage}
\end{figure}

Fig. \ref{fig:2:ospa} shows the OSPA distance for the current scenario. As
in the previous case study, the Consensus M$\delta $-GLMB filter outperforms
the Consensus CPHD filter. The same conclusion as in the previous case (High SNR) can be drawn for multiple consensus steps. 


\subsection{Low $P_{D}$}

\begin{figure}[h!]
	\begin{minipage}[t][][t]{\columnwidth}
		\centering
		\includegraphics[width=0.8\columnwidth]{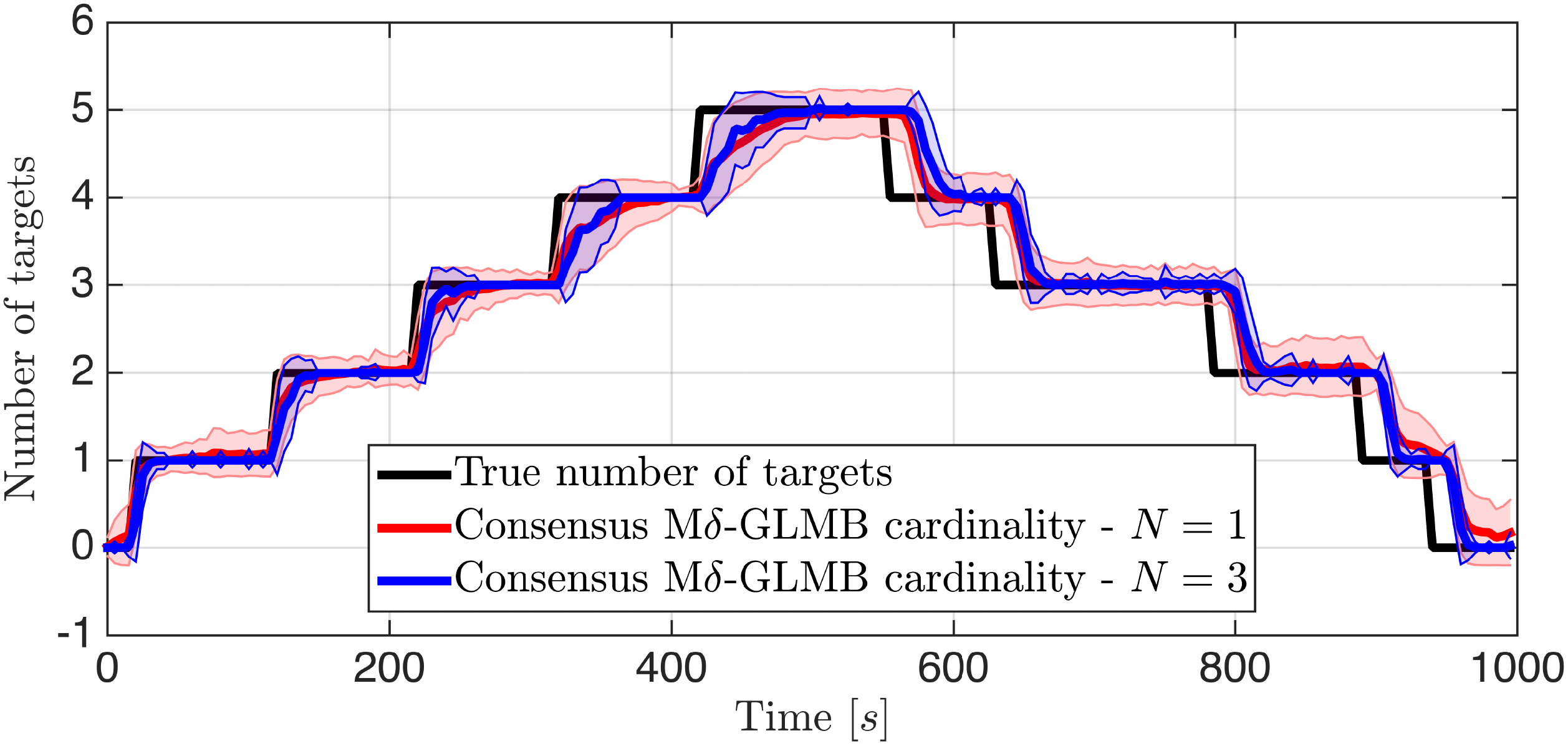}\vspace{-1em}
		\caption{Cardinality statistics for Consensus M$\delta$-GLMB tracker under low $P_{D}$.}
		\label{fig:3:cardCMDGLMB}
        \end{minipage}\vspace{1em}
	\begin{minipage}[t][][t]{\columnwidth}
		\centering
		\includegraphics[width=\columnwidth]{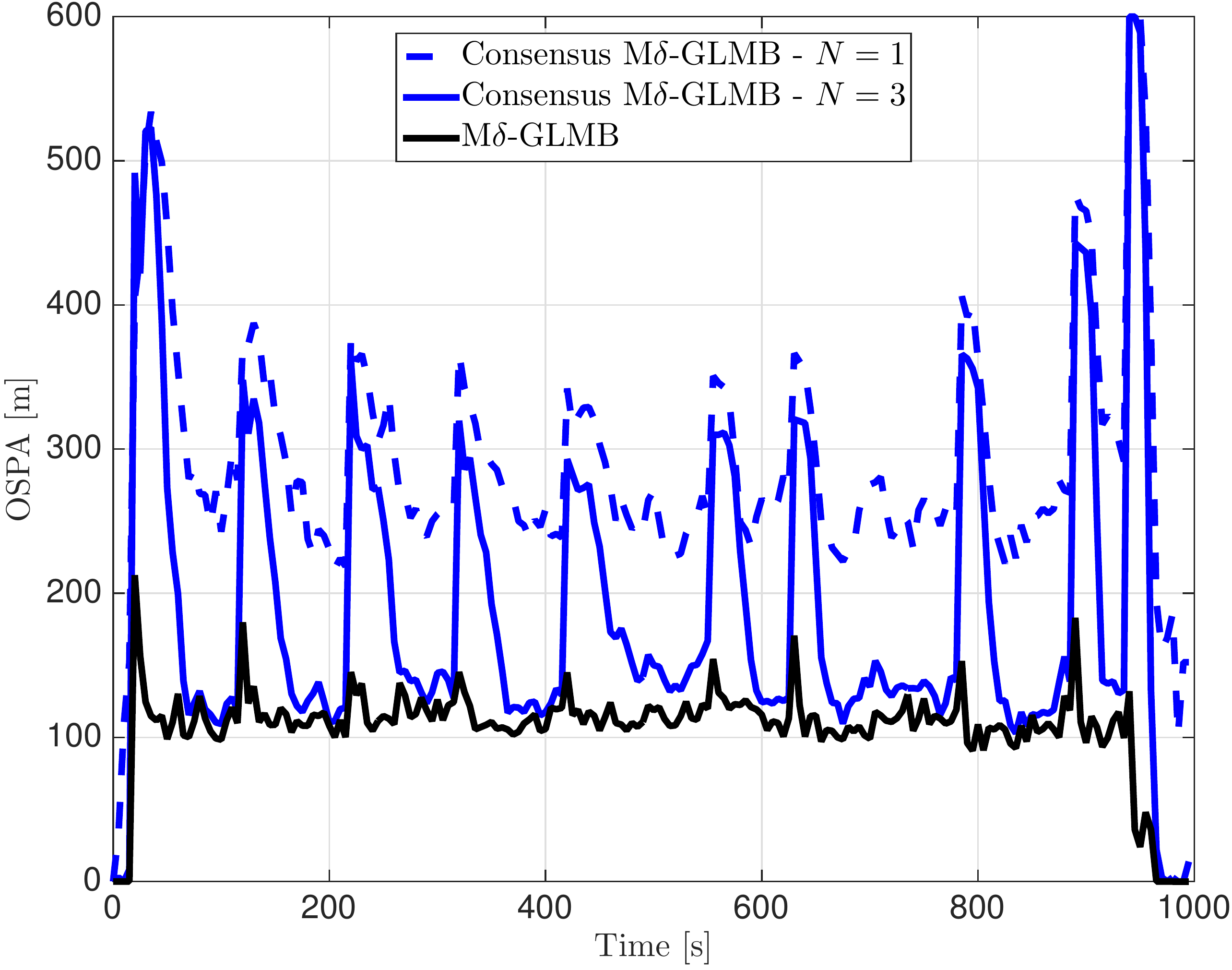}\vspace{-1em}
		\caption{OSPA distance ($c = 600 \, [m]$, $p = 2$) under low $P_{D}$.}
		\label{fig:3:ospa}
        \end{minipage}
\end{figure}

Fig. \ref{fig:3:cardCMDGLMB} displays the statistics (mean and standard
deviation) of the estimated number of objects obtained with the Consensus M$%
\delta $-GLMB.
It is worth noting that in this very challenging scenario with
$P_{D}=0.7$, the only working distributed algorithm is indeed the Consensus
M$\delta $-GLMB filter, and that it exhibits good performance in terms of
the average number of estimated objects. Fig. \ref{fig:3:ospa} shows the OSPA distance for the current scenario.

The benefit of using multiple consensus steps is particularly stressed by this simulation setting.
 As it can be seen from Figs. \ref{fig:3:cardCMDGLMB} and \ref{fig:3:ospa}, there is a remarkable improvement in both cardinality and state estimation error. Further, once the peaks in the OSPA reduce, the error is comparable to  the (centralized) M$\delta$-GLMB filter.

\section{Conclusions}

\label{sec:conclusions} In this paper, we have presented fully distributed
multi-object tracking solutions over a sensor network using labeled RFSs.
Consensus algorithms have been developed for fully distributed and scalable
fusion of information collected from the multiple heterogeneous and
geographically dispersed sensors. The proposed consensus algorithms are
based on the notion of Kullback-Leibler averaging of the local multi-object
probability densities. Efficient Gaussian mixture implementations have been
successfully tested on realistic multi-object tracking scenarios. Possible
topics for future work are to consider sensors with different field-of-view
and to investigate distributed measurement-driven object initialization.

\section{Appendix A}

\label{apx:proof}
Proof of Proposition \ref{pro:mdglmb:fusion}:\newline
Let $\eta^{(L)}(\ell)\triangleq \int \prod_{i\in \mathcal{I}}(p^{\left( i\right) \!}\left(
x,\ell; L \right) )^{\omega^{(i)}}dx$, and note from the definitions of $%
\overline{w}\!\left( L \right)$ and $\overline{p}\!\left( \cdot, L \right)$
in Proposition \ref{pro:mdglmb:fusion} that
\begin{IEEEeqnarray}{rCl}
	\overline{w}\!\left( L \right) & = & \frac{\displaystyle\prod_{i\in\mathcal{I}}(w^{\left( i\right) \!}(L))^{\omega ^{(i)}\!}[\eta ^{(L)}]^{L}}{\displaystyle\sum_{J\in \mathcal{F}\!\left( \mathbb{L}\right) }\prod_{i\in \mathcal{I}}(w^{\left( i\right) }(J))^{\omega ^{(i)}}[\eta ^{(J)}]^{J}} \label{eq:mdglmbproof_w}\\
	\eta ^{(L)}(\ell) \, \overline{p}\!\left( x,\ell; L \right) & = & \prod_{i\in \mathcal{I}} p^{\left( i\right) }\!\left(x,\ell; L \right)  \label{eq:mdglmbproof_a}
\end{IEEEeqnarray}

Using the form (\ref{eq:mdglmb2}) for M$\delta $-GLMB densities we have
\begin{IEEEeqnarray}{rCl}
	\prod_{i\in \mathcal{I}}(\boldsymbol{\pi}^{(i)}\left( \mathbf{X}\right) )^{\omega ^{(i)}} &=& \Delta \!(\mathbf{X})\!\prod_{i\in \mathcal{I}}(w^{(i)}\!\left(\mathcal{L}\!\left( \mathbf{X}\right) \right))^{\omega^{(i)}\!}\prod_{i\in \mathcal{I}}\!\left( [p^{(i)}\!\left( \cdot \right)]^{\mathbf{X}}\!\right) ^{\!\omega ^{(i)}} \notag \\
	&=& \Delta \!(\mathbf{X})\!\prod_{i\in \mathcal{I}}(w^{(i)}\!\left(\mathcal{L}\!\left( \mathbf{X}\right) \right))^{\omega ^{(i)}\!}\!\left[\prod_{i\in \mathcal{I}}(p^{(i)}\!\left( \cdot \right))^{\omega ^{(i)}\!}\right] ^{\mathbf{X}}  \notag \\
	&=& \Delta \!(\mathbf{X})\!\prod_{i\in \mathcal{I}}(w^{(i)}\!\left(\mathcal{L}\!\left( \mathbf{X}\right) \right))^{\omega ^{(i)}\!}[\eta ^{(\mathcal{L}\!\left( \mathbf{X}\right) )}]^{\mathcal{L}\!\left( \mathbf{X}\right) }[\overline{p}\!\left( \cdot; \mathcal{L}\!\left( \mathbf{X}\right) \right)]^{\mathbf{X}}  \label{eq: mdglmbproof1} \\
	&=& \sum_{J\in \mathcal{F}\!\left( \mathbb{L}\right) }\prod_{i\in\mathcal{I}}(w^{(i)}\!\left( J\right))^{\omega ^{(i)}\!}[\eta^{(J)}]^{J}\Delta \!(\mathbf{X})\delta _{\!J}(\mathcal{L}\!\left( \mathbf{X}\right) )[\overline{p}\!\left( \cdot; J\right)]^{\mathbf{X}}\label{eq: mdglmbproof2} \IEEEeqnarraynumspace
\end{IEEEeqnarray}
where (\ref{eq: mdglmbproof1}) follows by substituting (\ref%
{eq:mdglmbproof_a}).

Integrating (\ref{eq: mdglmbproof2}), applying Lemma 3 of \cite[Section III.B%
]{vovo1}, and noting that $\int \overline{p}\!\left( \cdot, \ell, J \right)
dx=1$ gives
\begin{IEEEeqnarray}{rCl}
	\int \prod_{i\in \mathcal{I}}(\boldsymbol{\pi }^{(i)}\left( \mathbf{X}\right) )^{\omega ^{(i)}}\delta
\mathbf{X} &=& \sum_{J\in \mathcal{F}\!\left( \mathbb{L}\right)}\prod_{i\in \mathcal{I}}(w^{\left( i\right) }\!\left( J \right))^{\omega ^{(i)}}[\eta^{(J)}]^{J}\!\!\!\!\int \!\!\!\Delta \!(\mathbf{X})\delta _{J\!}(\mathcal{L}\!\left( \mathbf{X}\right) )[\overline{p}\!\left( \cdot; J \right)]^{\mathbf{X}}\delta \mathbf{X}  \notag \\
	&=& \sum_{J\in \mathcal{F}\!\left( \mathbb{L}\right)}\prod_{i\in \mathcal{I}}(w^{\left( i\right) }\!\left( J \right))^{\omega ^{(i)}\!}[\eta^{(J)}]^{J}\!\sum_{L\in \mathcal{F}\!\left( \mathbb{L}\right) }\!\!\delta_{J\!}(L)  \notag \\
	&=& \sum_{J\in \mathcal{F}\!\left( \mathbb{L}\right)}\prod_{i\in \mathcal{I}}(w^{\left( i\right) }\!\left( J \right))^{\omega ^{(i)}}[\eta^{(J)}]^{J} \, .\label{eq:mdglmbproof3}
\end{IEEEeqnarray}

Dividing (\ref{eq: mdglmbproof1}) by (\ref{eq:mdglmbproof3}), and using (\ref%
{eq:mdglmbproof_w}) yields
\begin{equation*}
\frac{\displaystyle\prod_{i\in \mathcal{I}}\left( \boldsymbol{\pi }%
^{(i)}\left( \mathbf{X}\right) \right) ^{\omega ^{(i)}}}{\displaystyle\int
\prod_{i\in \mathcal{I}}\left(\boldsymbol{\pi }^{(i)}\left( \mathbf{X}%
\right) \right) ^{\omega^{(i)}}\delta \mathbf{X}} = \Delta \!(\mathbf{X})%
\overline{w}\!\left( \mathcal{L}\!\left( \mathbf{X}\right) \right)[\overline{%
p}\!\left( \cdot; \mathcal{L}\!\left( \mathbf{X}\right) \right)]^{\mathbf{X}}
\end{equation*}
which is an M$\delta $-GLMB with parameter set $\{(\overline{w}%
\!\left(L\right), \overline{p}\!\left( \cdot; L \right))\}_{L\in \mathcal{F}%
\!\left(\mathbb{L}\right) }$. Finally, using the equivalence between the KLA
and the normalized geometric mean in Theorem 1 completes the proof.

\section{Appendix B}
\label{apx:proof2}

Proof of Proposition \ref{pro:lmb:fusion}:\newline
Let
\begin{IEEEeqnarray}{rCl}
	\eta (\ell ) &\triangleq &\int \prod_{i\in \mathcal{I}}(p^{(i)}\!\left( x,\ell\right) )^{\omega ^{(i)}}dx  \label{eq:lmbproof_eta} \\
	\widetilde{q}\!\left( \ell \right) &\triangleq &\prod_{i\in \mathcal{I}}(1 - r^{(i)}\!\left( \ell \right))^{\omega ^{(i)}}  \label{eq:lmbproof_q} \\
	\widetilde{r}\!\left( \ell \right) &\triangleq &\eta (\ell )\prod_{i\in \mathcal{I}}(r^{(i)}\!\left( \ell \right))^{\omega ^{(i)}}  \label{eq:lmbproof_r}
\end{IEEEeqnarray}
Note from the definitions of $\overline{r}(\ell)$ and $\overline{p}(x ,\ell )
$ in Proposition \ref{pro:lmb:fusion} that
\begin{IEEEeqnarray}{rCl}
	\eta (\ell ) \, \overline{p}(x,\ell ) & = & \prod_{i\in \mathcal{I}} ( p^{(i)}\left( x,\ell \right) )^{\!\omega ^{(i)}\!} \, , \label{eq:lmbproof_a} \\
	\overline{r}\!\left( \ell \right) & = & \dfrac{\widetilde{r}\!\left( \ell \right)}{\widetilde{q}\!\left( \ell \right)+\widetilde{r}\!\left( \ell \right)}\,,1-\overline{r}\!\left( \ell \right)=\dfrac{\widetilde{q}\!\left( \ell \right)}{\widetilde{q}\!\left( \ell \right)+\widetilde{r}\!\left( \ell \right)} \, .  \label{eq:lmbproof_b}
\end{IEEEeqnarray}

Using the form (\ref{eq:LMB2Convert}) for LMB densities we have
\begin{IEEEeqnarray}{rCl}
	\prod_{i\in \mathcal{I}}(\boldsymbol{\pi }^{(i)}\left( \mathbf{X}\right) )^{\omega^{(i)}} &=& \Delta \!(\mathbf{X})\!\!\prod_{i\in \mathcal{I}}\!\!\left( [1\!-\!r^{(i)}]^{\mathbb{L}\backslash \!\mathcal{L\!(}\mathbf{X})}[1_{\mathbb{L}} \, r^{(i)}]^{\mathcal{L\!(}\mathbf{X})\!}\right) ^{\!\omega ^{(i)}}\!\!\prod_{i\in \mathcal{I}}\![(p^{(i)})^{\mathbf{X}}]^{\omega ^{(i)}}  \notag \\
	&=& \Delta \!(\mathbf{X})\!\!\prod_{i\in \mathcal{I}}\!\!\left( [1\!-\!r^{(i)}]^{\mathbb{L}\backslash \!\mathcal{L\!(}\mathbf{X})}[1_{\mathbb{L}} \, r^{(i)}]^{\mathcal{L\!(}\mathbf{X})\!}\right) ^{\!\omega ^{(i)}\!}\!\left[ \prod_{i\in \mathcal{I}}\!(p^{(i)})^{\omega ^{(i)}\!}\right] ^{\mathbf{X}}  \notag \\
	&=& \Delta \!(\mathbf{X})\!\!\left[ \prod_{i\in \mathcal{I}}\!(1\!-\!r^{(i)})^{\omega ^{(i)\!}\!}\right] ^{\!\mathbb{L}\backslash \!\mathcal{L\!(}\mathbf{X})}\!\!\left[ \!1_{\mathbb{L}}\,\!\prod_{i\in \mathcal{I}}\!(r^{(i)})^{\omega ^{(i)\!}\!}\right] ^{\!\mathcal{L\!(}\mathbf{X})}\!\!\!\eta ^{\mathcal{L\!(}\mathbf{X})}\overline{p}^{\mathbf{X}}  \notag \\
	&=& \Delta \!(\mathbf{X})\left[ \widetilde{q}\right] ^{\mathbb{L}\backslash \!\mathcal{L(}\mathbf{X})}\!\left[ 1_{\mathbb{L}}\left( \eta(\cdot) \prod_{i\in \mathcal{I}}\!(r^{(i)})^{\omega^{(i)}\!}\right) \right] ^{\!\mathcal{L\!(}\mathbf{X})}\!\overline{p}^{\mathbf{X}}  \notag \\
	&=& \Delta \!(\mathbf{X})\left[ \widetilde{q}\right] ^{\mathbb{L}\backslash \!\mathcal{L(}\mathbf{X})}\left[ 1_{\mathbb{L}}\,\widetilde{r}\right] ^{\mathcal{L(}\mathbf{X})}\!\overline{p}^{\mathbf{X}}\!\!  \label{eq:lmbproof1}
\end{IEEEeqnarray}
where the substitutions (\ref{eq:lmbproof_a}), (\ref{eq:lmbproof_q}) and (%
\ref{eq:lmbproof_r}) have been performed.

Integrating (\ref{eq:lmbproof1}), applying Lemma 3 of \cite[Section III.B]%
{vovo1} and noting that $\int \overline{p}\!\left( x,\cdot \right) \!dx=1$
gives
\begin{IEEEeqnarray}{rCl}
	\int \!\prod_{i\in \mathcal{I}}(\boldsymbol{\pi }^{(i)\!}\left( \mathbf{X}\right) )^{\omega ^{(i)}}\!\delta \mathbf{X} & = & \sum_{L\in\mathcal{F}\!\left( \mathbb{L}\right) }\left[ \widetilde{q} \right]^{\mathbb{L}\backslash L}\left[ \widetilde{r} \right] ^{L}  \notag\\
	& = & [\widetilde{q}+\widetilde{r}]^{\mathbb{L}}  \label{eq:lmbproof2}
\end{IEEEeqnarray}
where in the last step we applied the Binomial Theorem \cite{math}
\begin{equation*}
\sum_{L\subseteq \mathbb{L}}g^{\mathbb{L}\backslash L}f^{L} = [g+f]^{\mathbb{%
L}}.
\end{equation*}

Dividing (\ref{eq:lmbproof1}) by (\ref{eq:lmbproof2}), and using (\ref%
{eq:lmbproof_b}) yields
\begin{IEEEeqnarray*}{rCl}
	\frac{\displaystyle\prod_{i\in \mathcal{I}}\left(\boldsymbol{\pi }^{(i)}\left( \mathbf{X}\right) \right) ^{\omega ^{(i)}}}{\displaystyle\int \!\prod_{i\in \mathcal{I}}\left( \boldsymbol{\pi }^{(i)}\left( \mathbf{X}\right) \right) ^{\omega ^{(i)}}\!\!\delta \mathbf{X}} &=& \Delta (\mathbf{X})\frac{\left( \widetilde{q}\right)^{\mathbb{L}\backslash \!\mathcal{L}\!(\mathbf{X})}\left( 1_{\mathbb{L}} \, \widetilde{r}\right)^{\mathcal{L(}\mathbf{X})}}{(\widetilde{q}+\widetilde{r})^{\mathbb{L}}}\overline{p}^{\mathbf{X}}\\
	&=& \Delta (\mathbf{X})\frac{\left( \widetilde{q}\right) ^{\mathbb{L}\backslash \!\mathcal{L\!(}\mathbf{X})}\left( 1_{\mathbb{L}} \, \widetilde{r}\right) ^{\mathcal{L(}\mathbf{X})}}{\left( \widetilde{q}+\widetilde{r} \right)^{\mathbb{L}\backslash \!\mathcal{L\!(}\mathbf{X})} \left( \widetilde{q}+\widetilde{r} \right)^{\mathcal{L\!(}\mathbf{X})}}\overline{p}^{\mathbf{X}} \\
	&=& \Delta \!(\mathbf{X})\left( \dfrac{\widetilde{q}}{\widetilde{q}+\widetilde{r}}\right) ^{\mathbb{L}\backslash \!\mathcal{L\!(}\mathbf{X})}\left( 1_{\mathbb{L}} \, \dfrac{\widetilde{r}}{\widetilde{q}+\widetilde{r}}\right) ^{\mathcal{L\!(}\mathbf{X})}\overline{p}^{\mathbf{X}} \\
	&=& \Delta \!(\mathbf{X})\left( 1-\overline{r}\right) ^{\mathbb{L}\backslash \!\mathcal{L\!(}\mathbf{X})}\left( 1_{\mathbb{L}} \, \overline{r}\right) ^{\mathcal{L(}\mathbf{X)}}\overline{p}^{\mathbf{X}}
\end{IEEEeqnarray*}
which is an LMB with parameter set $\{(\overline{r}\!\left( \ell \right),
\overline{p}\!\left( \ell \right))\}_{\ell \in \mathcal{F}\!\left(\mathbb{L}%
\right) }$. Finally, using the equivalence between the KLA and the
normalized geometric mean in Theorem \ref{thm:kla} completes the proof.

\end{document}